%% file: Manuscript_JJohansson.tex
\documentclass[useAMS,usenatbib]{mn2e}

\usepackage[dvips]{graphicx}
\usepackage[fleqn]{amsmath}

\title[New Light on the Origin of Type Ia Supernovae]{Diffuse Gas in Galaxies Sheds New Light on the Origin of Type Ia Supernovae}
\author[J. Johansson et al.]{Jonas Johansson$^{1}$\thanks{E-mail: jjohansson@mpa-garching.mpg.de},  Tyrone E. Woods$^{1}$, Marat Gilfanov$^{1,2}$, Marc Sarzi$^3$, 
\newauthor Yan-Mei Chen$^{4,5}$ and Kyuseok Oh$^6$\\
$^1$Max-Planck Institut f{\"u}r Astrophysik, Karl-Schwarzschild-Str. 1, D-85741 Garching, Germany\\
$^2$Space Research Institute, Profsoyuznaya 84/32, 117997 Moscow, Russia\\
$^3$Centre for Astrophysics Research, University of Hertfordshire, College Lane, Hatfield, Herts, AL10 9AB, UK\\
$^4$Department of Astronomy, Nanjing University, Nanjing 210093, China\\
$^5$Key Laboratory of Modern Astronomy and Astrophysics (Nanjing University), Ministry of Education, Nanjing 210093, China\\
$^6$Department of Astronomy, Yonsei University, Seoul 120-749, Republic of Korea\\
}

\begin{document}

\pagerange{\pageref{firstpage}--\pageref{lastpage}} \pubyear{2013}
\maketitle
\label{firstpage}
\begin{abstract} 
We measure the strength of He{ \sc ii }$\lambda$4686 nebular emission in passively evolving (``retired") galaxies, aiming to constrain their populations of hot accreting white dwarfs (WDs) in the context of the single degenerate (SD) scenario of Type Ia supernovae (SNe~Ia).  
In the SD scenario, as a WD  burns hydrogen-rich material accreted from a companion star, it becomes a powerful source of ionizing UV emission. 
If significant populations of such sources exist in galaxies,  strong emission in the recombination lines of He~{\sc ii} should be expected from the interstellar medium. 
To explore this conjecture, we select from the Sloan Digital Sky Survey $\sim11~500$ emission line galaxies with stellar ages $>1$~Gyr showing no signs of AGN activity and co-add their spectra in bins of stellar population age. 
For the first time, we detect He~{\sc ii}~$\lambda$4686 nebular emission in retired galaxies and find it to be significantly weaker than that expected in the SD scenario, especially  in the youngest age bin ($1-4$ Gyr) where the SN~Ia rate is the highest.  
Instead, the strength of the observed He~{\sc ii}~$\lambda$4686 nebular emission is consistent with post-asymptotic giant branch stars being the sole ionizing source in all age bins.
These results limit populations of accreting WDs with photospheric temperatures ($T_{\rm eff}$) in the range $\sim(1.5-6)\cdot 10^5$ K to the level at which they  can account for no  more than $\sim 5-10\%$ of the observed SN~Ia rate.  
Conversely, should all WD progenitors of SN~Ia go through the phase of steady nuclear burning with  $T_{\rm eff}\sim(1.5-6)\cdot 10^5$ K, they do not increase their mass by more than $\sim 0.03 ~M_\odot$ in this regime.\end{abstract}


\input{Sec1}

\input{Sec2}
\input{Sec3}
\input{Sec4}

\input{Sec5}
\input{Sec6}

\section*{ACKNOWLEDGMENTS}

We would like to thank an anonymous referee for useful comments that helped improve this paper. JJ would like to thank the Deutsche Forschungsgemeinschaft (DFG) for financial support. KO acknowledges support from the National Research Foundation of Korea (SRC Program No. 2010-0027910) and the DRC Grant of Korea Research Council of Fundamental Science and Technology (FY 2012).

{}

\label{lastpage}

\input{appendixA}

\end{document}

%% file: Sec1.tex
\section{Introduction}

The use of Type Ia supernovae (SNe~Ia) as standard candles has been key to the discovery of the accelerating expansion of the Universe \citep{riess98,perlmutter99}. 
However, the nature of their progenitors remains uncertain, with increasing evidence suggesting they may arise through multiple evolutionary channels. This possible diversity, and the null detections of the progenitor of any individual SN Ia \citep[e.g.][]{2011fe} have rendered population-based arguments a necessity. 

The single-degenerate (SD) scenario \citep{whelan73} has long been one of the favoured progenitor channels for SNe~Ia. In the classical picture, a carbon-oxygen white dwarf (WD) accretes mass from a co-orbiting binary companion, eventually exploding upon nearly reaching the critical Chandrasekhar mass ($\sim$1.4 M$_\odot$). Before this happens, steady nuclear burning of the accreted material occurs on the surface of the WD over an extended period of time ($\sim$10$^6$ yr) as soon as the mass transfer rate ($\dot{M}$) exceeds a few times $10^{-7}M_\odot/yr$ \citep{nomoto07}. For lower mass transfer rates, nuclear burning occurs only in unsteady flashes, expelling matter from the WD in nova outbursts \citep{prialnik95}. This renders it difficult for the WD to ever reach the Chandrasekhar mass, unless perhaps if accreting WDs undergo an extended evolutionary phase as recurrent novae binaries \citep{hachisu01}. In the latter case, however, the SD scenario comes into conflict with statistics of classical/recurrent nova in the Andromeda galaxy (Soraisam \& Gilfanov, in preparation).

Should all SNe Ia progenitors undergo a steady nuclear-burning phase during some fraction of the time prior to explosion, their total bolometric luminosity L$_{tot}$ can easily be evaluated. It is proportional to the rate of exploding SN Ia ($\dot{N}_{SNIa}$), and to the total amount of mass $\Delta$M$_{SB}$ processed through steady burning, according to \citep{gilfanov}

\begin{equation}
L_{tot}=\epsilon_H\chi\Delta M_{SB}\dot{N}_{SNIa},
\label{Eq:Lum}
\end{equation}

\noindent where $\epsilon_H$ and $\chi$ are the energy release per unit mass of hydrogen ($\sim$6$\cdot$10$^{18}$ erg/g) and the mass fraction of hydrogen ($\sim$0.72), respectively.  

During steady nuclear-burning, the WD emits blackbody-like radiation with an effective temperature peaking around  $T_{\rm eff}\sim (5-8)\cdot 10^5$ K, making it a prodigious source of soft X-ray emission \citep[$\sim$0.3--0.7 keV,][]{vdHeuvel92}. Given the observed rate of SNe Ia \citep*{totani08,maoz12,graur14}, there should be a large number of accreting sources at any given time according to the SD scenario, sufficient to produce an integrated soft X-ray emission that would be detectable by modern X-ray telescopes, in particular in galaxies lacking an absorbing interstellar medium. However, recent studies found very little total soft X-ray emission in gas-poor elliptical galaxies \citep{gilfanov}, and far too few sources \citep{distefano}, implying that no more than $\sim$5\% of the SNe~Ia in these systems formed according to the standard SD scenario.

Such a direct argument is very efficient in constraining populations of relatively high temperature white dwarfs accreting in the steady nuclear burning regime. 
It is, however, much less sensitive to lower temperature sources (below $T_{\rm eff}\sim 4\cdot 10^5$ K), which may arise in the so-called accretion wind state of  the SD scenario, wherein the photosphere of the WD inflates to dimensions significantly exceeding the white dwarf radius.
This would shift the peak of
the radiation from the soft X-ray to the extreme UV regime, with
temperatures roughly around $1-2\cdot10^5$ K \citep*{hachisu10}. 
The total column density towards any such source would easily obscure it, preventing the direct detection of this UV emission. However, its signature will be seen as a hard ionizing source.

In fact, with harder UV spectra than young O-stars, old
horizontal-branch stars or post-asymptotic giant branch stars (pAGBs), the
presence of a SD-progenitor population will dramatically change
the character of the ionizing background of a galaxy, in particular in
the absence of a central active nucleus or for regions sufficiently
far from one. 
In turn, the hardness of the ionising background could be gauged in
the presence of a diffuse ionised-gas medium, by measuring the
strength of nebular emission that can only be powered by a far-UV
background, e.g. the He~{\sc ii} recombination line at
4686 \AA\ (He~{\sc ii}~$\lambda$4686). The strength of this line is indeed predicted to be
considerably boosted should SNe Ia originate from hot and
luminous WD progenitors compared to a situation where the ionizing
background is dominated by old UV-bright stars \citep{woods}.

Hence, in this paper, the aim is to examine whether a substantial
population of steadily-accreting WDs with black-body temperatures
around a few times $10^5$ K is consistent with measurements for the strength
of He~{\sc ii}~$\lambda$4686 emission from the diffuse ionized-gas of
galaxies, which we will use as a gauge for the hardness of the
ionizing background. 
In particular, we focus on galaxies with $>$1~Gyr old
passively-evolving stellar populations (also referred to as ``retired
galaxies'') with no sign of active galactic nuclei (AGN). Such a
selection excludes the contribution of young and hot stars or of an
accreting supermassive black hole to the ionizing background, leaving
us to consider only the presence of old UV-bright stars such as pAGB
stars in addition to the possible impact of accreting WDs. 
Retired galaxies have been shown to often host substantial
amounts of neutral gas largely confined to a disk-like planar
distribution within their host galaxy \citep{ATLAS3D}. This neutral
gas is accompanied by extended regions of lowly-ionized gas
(e.g., Sarzi et al. 2006), the emission of which has long been thought
to be powered by a diffuse ionizing background supplied by stellar
populations (Binette et al. 1994).

Retired galaxies are therefore an ideal laboratory to check whether
a substantial accreting WD population of SN Ia progenitors exists. 
Yet, even though the presence of such hot sources would substantially
boost the observed He~{\sc ii}~$\lambda$4686 emission from the
interstellar medium of retired galaxies, the recombination lines remain
intrinsically faint and their detection very challenging. 
For instance, in the case where pAGB stars are the only sources of He~{\sc
ii}~-ionizing photons, the strength of the He~{\sc ii}~$\lambda$4686-line
is predicted to require a
signal-to-noise (S/N) ratio in the stellar continuum $>$500 to be detected 
\citep{woods}. This high-quality requirement contributes not only to
explain why to date there is no reported detection of He~{\sc
ii}-emission in retired galaxies, but also sets the bar for the
accuracy needed for our experiment if we wish to recognize an increase
in the He~{\sc ii} strength due to the presence of accreting WDs. 

To achieve the required data quality, we will draw from the great wealth of
optical spectra of nearby galaxies available in the Sloan
Digital Sky Survey \citep[SDSS,][]{york00} and co-add a large number
of spectra of passively evolving systems with no sign of recent star
formation or AGN activity, but for which we can detect the same
kind of low-ionization nebular emission that is generally observed in
nearby early-type galaxies. 
Such an emission is indeed always extended and has been shown to be
consistent with the weak emission found in the SDSS spectra of retired
galaxies, which cover most of the optical regions of these objects
(Sarzi et al. 2010).

This paper is organized as follows. The data sample selection is
described in Section~\ref{data}. The technique used for stacking
galaxy spectra and the method for estimating emission lines are
described in Section~\ref{stacking}. The derived He~{\sc
  ii}~$\lambda$4686 emission is presented Section~\ref{results} and
compared to modelled predictions for the SD scenario in
Section~\ref{disc}. Concluding remarks are given in
Section~\ref{conc}.

%% file: Sec2.tex
\section{Data sample}
\label{data}

The sample used in this study is selected from the SDSS-II data release 7 \citep[DR7][]{DR7}.
The SDSS has obtained imaging \citep[\textit{ugriz};][]{fukugita96} and spectroscopy for more than 930~000 galaxies.
This was achieved with the multi-object spectrograph \citep{smee12}, fed with light from fibres with a fixed diameter of 3", and wide-field CCD camera \citep{gunn98} on the SDSS 2.5 m telescope \citep{york00,gunn06} at the Apache Point Observatory.  Starting from the full DR7 sample we apply a number of selection criteria described below. The numbers of the samples after applying each cut are presented in Table~\ref{tab:sample}.

\subsection{Redshift and WHAN selection}
\label{data:redshift}

For the stacking procedure we want to avoid galaxies for which the fixed SDSS aperture only covers the nuclear regions. Hence, we apply a lower redshift limit of $z>0.04$, such that all spectra cover the galaxy stellar light within a diameter $>\sim2.5$ kpc. While the wavelength coverage of the SDSS spectra allows for H$\alpha$ to be detected up to z~$\sim0.4$, we do also apply an upper redshift limit of $z<0.1$, above which the signal-to-noise (S/N) of the SDSS spectra decreases fast. There are $\sim$306~000 SDSS DR7 galaxies for the adopted redshift range of $0.04<z<0.1$ (``Redshift cut'' in Table~\ref{tab:sample}).

\begin{table}
\center
\caption{The corresponding number of galaxies left after applying each of the different sample cuts, starting from the full SDSS DR7 catalogue.}
\label{tab:sample}
\begin{tabular}{lr}
\hline
\bf Selection   &  \multicolumn{1}{c}{\bf N} \\
\hline
DR7      & 927~552 \\
Redshift cut &  305~598 \\
WHAN cut & 13 875 \\
RSF cut {\sc I} (PCA) & 11 593 \\
RSF cut {\sc II} (NUV-$r$) & 4 061 \\
\hline
\end{tabular}
\end{table}

\begin{figure}
\centering
\includegraphics[clip=true,trim=0.2cm 2cm 1cm 2cm,scale=0.5,angle=90]{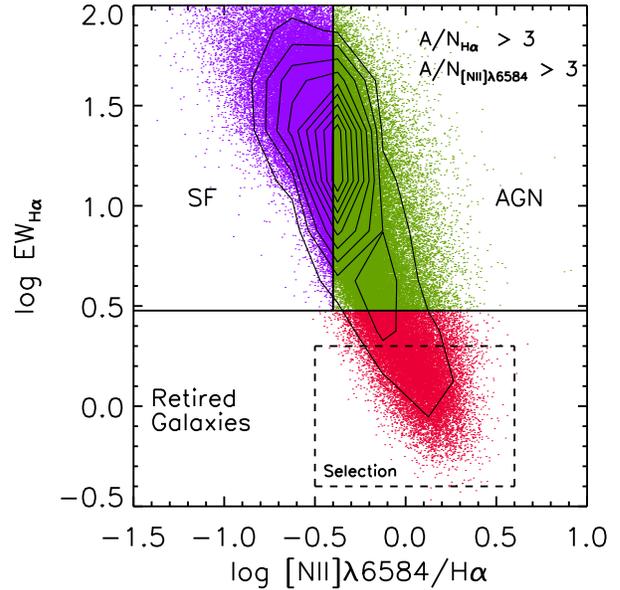}\\
\caption{The WHAN-diagram \citep{fernandes11}, [N~{\sc ii}~]$\lambda$6584/H$\alpha$ line ratio versus the
  strength of H$\alpha$ (quantified by its equivalent width), for SDSS
  galaxies with detected H$\alpha$ and [N~{\sc ii}~]$\lambda$6584 emission, i.e. for
  which the A/N~$>3$ for both lines. The regions defined by the solid lines
  include galaxies with ionizing UV-radiation dominated by young stars
  in star-forming (SF) regions (top left), an AGN (top right) or UV-bright
  sources in old stellar populations (bottom). The dashed box
  represents our selection of old, so-called retired galaxies.}
\label{selection}
\end{figure}

Most critically, we want to select gas rich galaxies where the ionizing photon field is not powered by young OB-stars or accreting black holes. 
For this purpose we use the combination of H$\alpha$ line flux and
H$\alpha$/[N~{\sc ii}~]$\lambda$6584 line ratio, the so called WHAN diagram \citep{fernandes11}, 
to discriminate between objects where the nebular emission is powered by star-formation, an active galactic nucleus (AGN) or old and UV-bright stellar sources such as pAGB stars. Galaxies with detected H$\alpha$ and [N~{\sc ii}~]$\lambda$6584 emission were selected from the OSSY value-added SDSS catalogue
\citep{oh11}. In this
process objects with robust emission-line fits of H$\alpha$ and [N~{\sc ii}~]$\lambda$6584,
i.e. with an amplitude-over-noise (A/N) ratio $>3$ for both lines, were selected. Furthermore, 
objects with telluric
emission contaminating these lines were excluded \citep{oh11}.
The WHAN-diagram for the selection of galaxies with detected H$\alpha$ and [N~{\sc ii}~]$\lambda$6584 lines is shown in Fig.~\ref{selection}. Region ``SF" covers star-forming galaxies, region ``AGN" galaxies hosting AGNs and region ``Retired Galaxies" where old UV-bright stellar sources alone should
suffice to ionize the interstellar medium \citep{fernandes11}. The selection-limit of the latter region is $\log$ EW$_{\rm H\alpha} <0.48$. We adopt a more conservative limit of $\log\textrm{EW}_{\textrm{H}\alpha}=0.3$ to minimize the AGN contamination. 
This selection criterion, indicated by the dashed box in Fig.~\ref{selection}, produces a sample of 13~875 galaxies (``WHAN cut'' in Table~\ref{tab:sample}). 

\begin{figure}
\centering
\includegraphics[clip=true,trim=0.7cm 1cm 1.5cm 4cm,scale=0.45,angle=90]{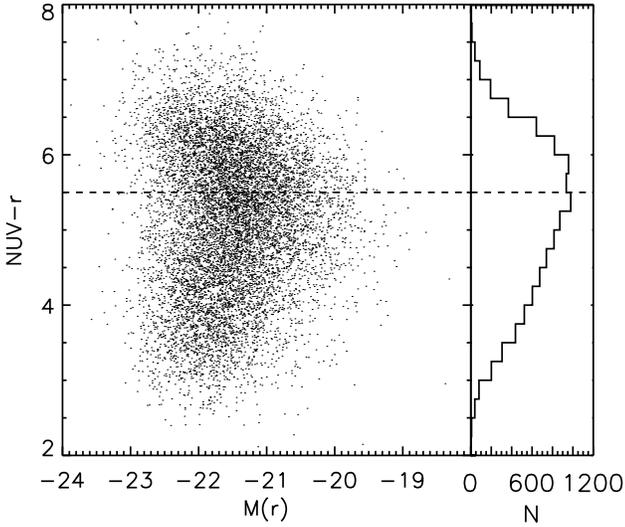}\\
\caption{Left panel: NUV-$r$  colour as a function of absolute $r$ band magnitude M($r$) for the 10~406  WHAN selected galaxies with available GALEX NUV magnitudes. The dashed line represents the limit from \citet{kaviraj07}, for which galaxies falling in the region below have experienced star formation in the last Gyr. Right panel: Distribution of NUV-$r$ colours for the same sample selection as above.}
\label{NUV-r}
\end{figure}

\subsection{Selecting galaxies without star formation in the last Gyr}
\label{data:sf}

In the last selection step, we discard galaxies hosting
a significant fraction of stars younger than 1~Gyr. For this purpose we use the
principal component analysis (PCA) technique of \citet{chen12} that is based on the work of \citet{wild07}.
The PCA method reproduces galaxy spectra by a set of orthogonal principal components.
These components represent spectral features including the 4000~\AA~break and Balmer absorption lines that are very sensitive to the presence of young stars.
Important for this work, the PCA technique of \citet{chen12} exploits these features to trace the fraction of stars formed in the last Gyr. 
After running a PCA on the individual galaxy spectra, we
exclude objects where more than one per cent of the stellar mass has formed in a recent star formation (RSF) episode in the last
Gyr, resulting in a final sample of 11~593 galaxies (``RSF cut {\sc I} (PCA)'' in Table~\ref{tab:sample}).
 
Complementary to the PCA technique, we separately exclude galaxies with recent star formation by combining  near ultra-violet (NUV) photometry with SDSS $r$ band magnitudes. For this purpose, we cross-match the selected SDSS sample, i.e. WHAN cut (see Table~\ref{tab:sample}), with the galaxy evolution explorer (GALEX) data base \citep{martin05} to retrieve NUV magnitudes for the individual galaxies. We find available GALEX photometry for 75~per cent of the selected galaxies, i.e. for 10~406 out of the 13~875 objects (see Section~\ref{data:redshift}) 
in our sample. 
Objects with any star formation in the last Gyr typically have NUV-$r$ colours $<5.5$ \citep{kaviraj07}. The 10~406 galaxies with available NUV magnitudes are shown in Fig.~\ref{NUV-r} for NUV-$r$ colour versus absolute $r$ band magnitude M($r$) (left hand panel), where the dashed line indicates the limit from \citet{kaviraj07}. The right hand panel shows the histogram of NUV-$r$ colour. This selection criterion is more conservative than using the PCA approach and leaves a final sample of 4~061 galaxies (``RSF cut {\sc II} (NUV-$r$)'' in Table~\ref{tab:sample}). This smaller sample leaves less freedom in the co-addition and analysis of stacked spectra, but it is a good complement to test and confirm the results of the PCA selected sample.

\subsection{Stellar age estimates and age grouping}
\label{data:ages}

In order to better constrain the predictions concerning the strength
of the He~{\sc ii}~$\lambda$4686 line within the single-degenerate progenitor model for
SN~Ia, we divided our final galaxy samples (``RSF cut {\sc I} (PCA)''  and ``RSF cut {\sc II} (NUV-$r$)'' in Table~\ref{tab:sample}) into several bins of average stellar population age. 
The ages are derived, together with stellar metallicity and element abundance ratios (O/Fe, Mg/Fe, C/Fe, N/Fe, Ca/Fe and Ti/Fe), using the technique from \citet{johansson12} which is based on absorption line indices. This technique was developed to break the degeneracy, exhibited by the integrated light of stellar populations, of stellar age with both stellar metallicity and element abundance ratios. Moreover, absorption line indices are not prone to extinction uncertainties due to their narrow wavelength definitions. The reader is referred to \citet{johansson12} for details of the method, while only a brief description is given here.

The main working algorithm of the method is the fitting of observed and modelled absorption line indices using a $\chi^2$-minimization routine. The indices are defined for 25 prominent absorption features in the optical, known as the Lick indices \citep{worthey94}. The observed Lick indices are taken from the OSSY data base. Following standard procedures, these indices were measured on emission line corrected absorption spectra downgraded to the Lick/IDS resolution \citep{worthey97}, and finally corrected for velocity dispersion broadening. The $\chi^2$-minimization routine adjusts the stellar population models of Lick indices from \citet*[][TMJ]{TMJ11} in order to match the observed values. This is done through a number of steps in order to carefully account for the sensitivity to the full range of element abundance ratios of the Lick indices. At each step a new set of models is used that is a perturbation of the previous best-fitting model. The best-fitting model at the final step, i.e. when the $\chi^2$  is not improved by more than 1 per cent, determines the stellar population parameters of the galaxy spectrum examined.

\begin{figure}
\centering
\includegraphics[clip=true,trim=0.7cm 2cm 1.5cm 2.5cm,scale=0.5,angle=90]{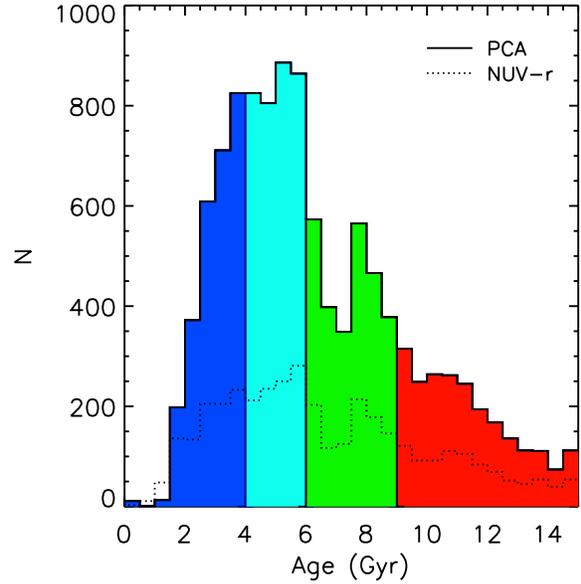}\\
\caption{Distribution for the luminosity-weighted mean stellar age of
  our selected sample of retired galaxies. The coloured
  sub-histograms represent our four adopted age bins, which contain a
  similar number of galaxies. The solid and dashed histograms represent the samples where galaxies with star formation in the last Gyr have been discarded using the PCA and NUV-$r$ criteria (see text), respectively, as indicated by the labels in the upper right corner.}
\label{age_distr}
\end{figure}

\begin{table}
\center
\caption{ Salient features of each of the co-added spectra analysed in
  this work. The columns give the age range of the galaxies in each group (col.~1), method for discarding galaxies with stars formed in the last Gyr (col.~2), number
  of spectra in each group (col.~3) and resulting S/N for the stacked
  spectra (col.~4) computed using the propagation of the
formal SDSS uncertainties (see Section~\ref{stacking:technique}). }
  \label{table:stacks}
\begin{tabular}{ccrr}
\hline
 \multicolumn{2}{c}{\bf Stack }  & \bf N~~ & \bf S/N \\
 \hline
All ages  & PCA &11593 & 1751 \\
$<$4 Gyr & PCA & 2740  &  882 \\ 
4-6  Gyr & PCA & 3380  & 1044 \\
6-9  Gyr & PCA & 2729  & 1000 \\
$>$9 Gyr & PCA & 2744  & 1018 \\ 
$<$6 Gyr & NUV-$r$ & 1953 & 829 \\
$>$6 Gyr & NUV-$r$ & 2108 & 951 \\
\hline
\end{tabular}
\end{table}

This routine is run for all individual galaxy spectra included in the final selections (``RSF cut {\sc I} (PCA)'  and `RSF cut {\sc II} (NUV-$r$)'  in Table~\ref{tab:sample}). The age distributions for the selections are shown in Fig.~\ref{age_distr}, where the solid histogram shows the distribution for the PCA selection and the dotted histogram shows the NUV-$r$ selected analogue. The former distribution is divided in bins containing at least $\sim2000$ objects, as shown by the coloured sub-histograms. This is the number needed to reach a desired S/N of $\sim500$ of the stacked spectra for the detection of the He~{\sc ii}~$\lambda$4686 lines
with as low predicted equivalent width values as
$\sim0.05$ \AA, according to \citet{woods}, estimated using the mean S/N of the SDSS spectra of $\sim$15.
Hence, the sample is divided into four age bins ($<4$ Gyr, $4-6$ Gyr, $6-9$ Gyr and $>9$ Gyr), each with a
number of objects greater than 2000. These numbers are presented in Table~\ref{table:stacks} for each age bin together with the resulting S/N after stacking all individual spectra (see Section~\ref{stacking:technique}). 
The more conservative cut using the NUV-$r$ colour allows for a division into two age bins ($<6$ Gyr and $>6$ Gyr), in order to have $\sim2000$ galaxies in each bin. 

%% file: Sec3.tex
\section{Co-addition and Analysis of the resulting Stacked Spectra}
\label{stacking}

\subsection{Stacking technique}
\label{stacking:technique}

Prior to stacking the spectra of the different sample selections, a few corrections are applied to the individual spectra:

\begin{enumerate}
\item The additional flux calibration of \citet{yan11} is applied to correct for small-scale fluctuations.
\item The spectra are de-reddened for Galactic extinction following the extinction curve of diffuse gas from \citet{odonnell94} for an R$_v$ value of 3.1, and using the Galactic E(B--V) values from the maps of \citet*{schlegel98}.
\item The spectra are brought to rest-frame wavelengths and to a common wavelength grid.
\end{enumerate}

After applying these corrections, the spectra of each sample selection are co-added in emitted flux to produce high S/N stacked spectra. We have also co-added the spectra in emitted luminosity and after normalizing them using the average flux in a wavelength window of $\pm$250 \AA\ around 5000 \AA. The results from using the different stacking techniques are very similar, i.e. for the studied emission line ratio He~{\sc ii}~$\lambda$4686/H$\beta$ (see Section~\ref{results}) the difference is less than the 1$\sigma$-errors. Hence, the results are not biased by the adopted stacking technique. During the stacking process, the statistical error arrays associated with each SDSS spectrum are co-added in quadrature. Computed using this propagation of the formal SDSS uncertainties, the mean S/N around 5000 \AA\ ($\pm250$ \AA) for all sub-samples are presented in Table~\ref{table:stacks} (col.~4)  and vary between $\sim830$ and $\sim1040$ for the different age bins and reach a value of $\sim1750$ for stacking all galaxies selected  with the PCA criterion (see Section~\ref{data:sf}).

It should be noted that redshift uncertainties have not been accounted for in the stacking process. Instead, we have verified, through Monte Carlo simulations, that including these uncertainties, the flux density errors of the stacked spectra increase by $\sim$10 per cent only.  We also note that the S/N values in Table~\ref{table:stacks} are presented for illustrative purposes and should be considered approximations for characterizing the quality of the stacked spectra. Using the residual noise level of the fits (see Section~\ref{stacking:EMfitting}) as an additional way of estimating the S/N ratios, we find similar but typically lower values by $\sim10-30$~\% for the different age bins.

\begin{figure}
\centering
\includegraphics[clip=true,trim=-0.5cm 1cm 1.5cm -0.30cm,scale=0.37,angle=90]{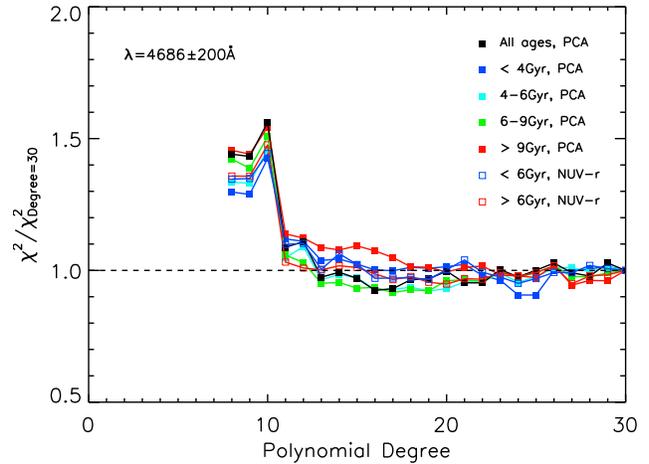}\\
\caption{The $\chi^2$ value derived in a wavelength window $\pm200$ \AA\ around 4686 \AA\ as a function of the polynomial degree used for adjusting the stellar templates in the SED fitting procedure (see text), normalized to the $\chi^2$ value using a polynomial degree of 30. The various colours represent the different age bins used for stacking spectra as given by the labels.}
\label{degree}
\end{figure}

\begin{figure*}
\centering
\includegraphics[clip=true,trim=0cm 0cm 0cm 0cm,scale=0.58,angle=90]{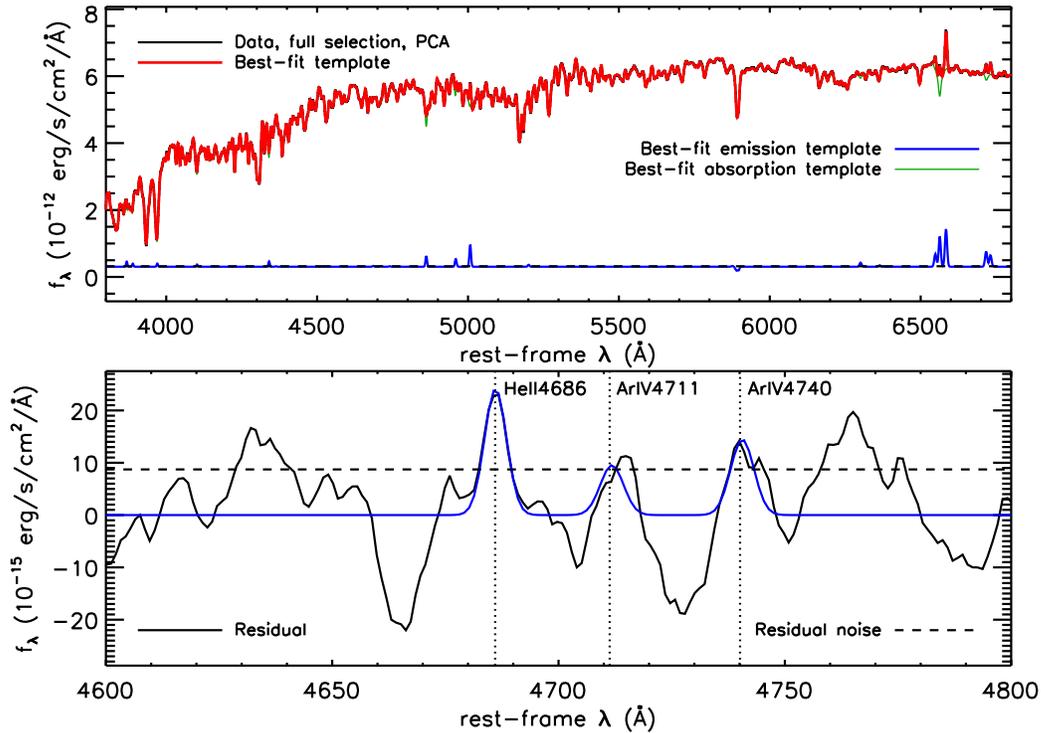}\\
\caption{Data and model of stacked spectrum. Upper panel: total spectrum obtained by co-adding the spectra
  of all galaxies in our sample of quiescent objects (11593
  galaxies) and selected using the PCA criterion (see Section~\ref{data}), together with our best-fitting model (red line). The
  stellar component of this model is shown by the thin green line,
  whereas our best-fitting nebular spectrum is shown in blue. Lower
  panel: difference between the stacked spectrum and best-fitting
  stellar model around the position of the He~{\sc ii}~$\lambda$4686 line (black solid line),
  together with our best-fitting emission-line model (blue line). The
  dashed horizontal line shows the level of noise in the residuals
  (standard deviation) of the overall fit to the data in a wavelength region $\pm$200 \AA\ around 4686 \AA. In the
  4600-4800 \AA\ wavelength range, beside He~{\sc ii}~$\lambda$4686, we also fit
  for the [Ar~{\sc iv}~] lines at 4711 and 4740 \AA. The position of all lines
  is indicated by dotted vertical lines, with labels for their
  name. In this case, the ratio between the Gaussian model amplitude
  and the level of noise in the residuals, A/N, which relates to the
  detection level of the lines, is 2.8 for the He~{\sc ii}~$\lambda$4686-line.}
\label{fit:HaLT0p4}
\end{figure*}

\subsection{Emission line fitting}
\label{stacking:EMfitting}

With the aim of measuring emission lines we adopt the fitting code {\sc gandalf} \citep{sarzi06}. A brief description of this code is given here, while the reader is referred to \citet{sarzi06} for a detailed description. The goal of {\sc gandalf} is the separation of emission and absorption spectra of composite galaxy spectra. In the first step, the emission lines of the wavelength range under consideration are masked and the {\sc ppxf}
code \citep{ppxf} is used for measuring kinematic broadening of the absorption spectrum with a set of best-fitting stellar templates. The emission line mask is then lifted and with the stellar kinematics fixed, {\sc gandalf} reassess  the stellar continuum by simultaneously fitting a number of Gaussian emission line templates consisting of both re-combination and forbidden lines.

As stellar templates we adopt the full medium-resolution
Isaac Newton telescope library \citep[MILES,][]{miles}. This library consists of
observed spectra for 985 solar neighbourhood stars. This is the stellar library with the largest parameter coverage in the literature and, hence, allows us to achieve a very precise fit
to the stellar continuum of the stacked spectra. The precision of the fit is significantly higher compared to using stellar population models as templates
\citep{capp07,sarzi10}. 

The MILES library has a fixed instrumental full width half maximum (FWHM) resolution $\Delta\lambda\sim2.5$ \AA\ \citep{beifiori11}, while the SDSS resolution is fixed at R~$\sim2000$ which corresponds to a FWHM $\Delta\lambda\sim2.5$ \AA\ at 5000 \AA . Even though the resolutions are close, the MILES resolution is  slightly higher at wavelengths longer than 5000
\AA\ and vice versa. Hence, we degrade both the SDSS and MILES spectra to match each other prior to running {\sc gandalf}.

We perform our {\sc gandalf} fitting in the wavelength window
between 3800 \AA\ and 6800 \AA, which includes many prominent absorption
features useful for constraining the best-fitting set of stellar
templates, as well as a wide range of important emission lines. This wavelength range
specifically covers the strong H$\alpha$ and
[N~{\sc ii}~]$\lambda$6584 lines, which are useful for constraining the position and
Gaussian width of weaker emission lines such as He~{\sc ii}~$\lambda$4686 and
H$\beta$. In particular, we tied the kinematics of all forbidden
lines, such as the [O~{\sc iii}~]$\lambda$$\lambda$4959,5007 doublet, to the
[N~{\sc ii}~]$\lambda$6584 kinematics, whereas all recombination lines were
tied to H$\alpha$.

\begin{table*}
\center
\caption{Age range and selection criterion of the galaxies in each group (col.~1-2), cols.~3-5 list the A/N ratio corresponding to
  our fit for the He~{\sc ii}~$\lambda$4686, H$\beta$ and [O~{\sc iii}~]$\lambda$5007
  lines, respectively, whereas cols.~6 and 7 list the values for the
  He~{\sc ii}~$\lambda$4686/H$\beta$ and [O~{\sc iii}~]$\lambda$5007/H$\beta$ line ratios. Col.~8 gives the derived average stellar population age.}
\label{AoN}
\begin{tabular}{ccccccccr@{}l}
\hline
 \multicolumn{2}{c}{\bf Stack } &  \multicolumn{3}{c}{\bf A/N} & \bf He~{\sc ii}~/H$\beta$  & \bf [O~{\sc iii}~]/H$\beta$  &  \multicolumn{3}{c}{\bf Age (Gyr)}  \\
&  &\bf He~{\sc ii}~ & \bf H$\beta$ & \bf [O~{\sc iii}~] &  &  \\
\hline
All        & PCA & 2.8 & 32.2 & 53.4 & 0.073$\pm$0.021 & 2.24$\pm$0.06 && 4&.7  \\ 
$<$4 Gyr    & PCA & 2.7 & 32.6 & 51.9 & 0.075$\pm$0.023 & 1.95$\pm$0.06 && 2&.8\\ 
4-6  Gyr    & PCA & 3.0 & 30.9 & 53.6 & 0.089$\pm$0.023 & 2.22$\pm$0.07 && 4&.9 \\ 
6-9  Gyr    & PCA & 2.6 & 28.0 & 47.8 & 0.087$\pm$0.026 & 2.31$\pm$0.08 && 6&.6 \\ 
$>$9 Gyr   & PCA & 2.6 & 23.8 & 39.9 & 0.103$\pm$0.029 & 2.32$\pm$0.09 && 10&.7 \\ 
$<$6 Gyr   & NUV-$r$ & 2.7 & 28.5 & 52.4 & 0.088$\pm$0.023 & 2.28$\pm$0.07 && 3&.0 \\ 
$>$6 Gyr   & NUV-$r$ & 2.6 & 23.8 & 40.1 & 0.101$\pm$0.026 & 2.31$\pm$0.09 && 8&.7\\ 
\hline
\end{tabular}
\end{table*}

When running {\sc ppxf} and {\sc gandalf} we adjust the
fit of the templates to the stellar continuum using multiplicative polynomials up to a
relatively high order, 15, capable of affecting the continuum shape
at a 200 \AA~level. 
Using a simple reddening law instead of
polynomials would lead to considerably worse fits, most likely owing
to residual uncertainties in the flux calibration of both object and
template spectra.
In Appendix~A we evaluate the effect of adjusting the stellar continuum using multiplicative polynomials and 
conclude that for the great part the polynomial correction accounts for the effects of reddening by dust, with higher order terms being needed for additional weak corrections on small scales (see Fig.~A8).

The polynomial degree was chosen to maximize the quality of the fit around the He~{\sc ii}~$\lambda$4686 ($\pm$200
\AA) and we find that the $\chi^2$ does not generally decrease for degrees greater than 15. 
This is shown in Fig.~\ref{degree} where the $\chi^2$ values, normalized to the $\chi^2$ value for a degree of 30, are presented as a function of polynomial degree. 
The $\chi^2$-ratio generally flattens out at a degree of 15, hence, there is no gain in using a higher degree for detecting the He~{\sc ii}~$\lambda$4686 line. From now on, only results for adopting a degree of 15 are presented and discussed.
Once an emission line is fit by {\sc gandalf}, we trace
its level of detection through the A/N ratio between the Gaussian
model peak amplitude and the level of noise in the residuals of the
global fit, which we measure within a $\pm$200~\AA-wide wavelength
window around the line of interest. Hence, from now on we refer noise to the residual noise, instead of the propagation of the formal SDSS uncertainties used as a measure of the noise in Section~\ref{stacking:technique}. Generally, a value of A/N greater
than 3 is considered sufficient to grant detection of a line for which the
kinematics has been held fixed \citep{sarzi06}. However, in
Section~\ref{results:HeII}, we further investigate the detection level
associated with any given A/N value in the case of the He~{\sc ii}~$\lambda$4686
line.

The ability to estimate the amplitude of the lines, which depends on the residual noise level, dominates the error of the line fluxes \citep{sarzi06}. 
Statistical errors are not the only contribution to the residual noise level, but systematic uncertainties, or template mismatch, will contribute as well. Hence, to properly estimate the amplitude errors the statistical errors are scaled to get a $\chi^2\sim1$ for the overall fit \citep{sarzi05}. 

As already mentioned, for the adopted wavelength range we fit a large number of emission lines. However, in this paper we focus on the detection of the He~{\sc ii}~$\lambda$4686 line.  In addition to this line, we will also use the results for H$\beta$ and [O~{\sc iii}~]$\lambda$5007 (see Section~\ref{disc}) in order to constrain the ionization state of the gas. All other detected optical lines will be discussed in a companion paper (Johansson et al., in preparation).

%% file: Sec4.tex
\section{Stacking Results}
\label{results}

In this section, we present the results for the lines He~{\sc ii}~$\lambda$4686, H$\beta$ and [O~{\sc iii}~]$\lambda$5007 from the SED fitting of the stacked spectra using {\sc gandalf}. In particular, we assess the confidence level of the detection of the weak He~{\sc ii}~$\lambda$4686 line. Furthermore, we derive the average stellar population age of the emission corrected stacked spectra.

\subsection{Emission lines}
\label{results:HeII}

The spectrum obtained when stacking all selected galaxies using the PCA criterion (see Section~\ref{data:sf}) is presented in Fig.~\ref{fit:HaLT0p4} (upper panel) together with our best-fitting model.  
The stacked spectrum is matched by the best-fit template to a very high precision and a range of prominent emission lines are apparent. 
Corresponding figures for the rest of the stacked spectra are presented in Appendix~A (Fig.~A1-A6), together with a zoom-in on the fit around He~{\sc ii}~$\lambda$4686, H$\beta$ and [O~{\sc iii}~]$\lambda$5007 (Fig.~A7). 
The lower panel of Fig.~\ref{fit:HaLT0p4} shows the difference between the stacked spectrum and best-fitting stellar template, together with the best-fitting emission-line model around the position of He~{\sc ii}~$\lambda$4686. A positive feature is clearly visible at the location of the He~{\sc ii}~$\lambda$4686 line, which is well matched while adopting the same Gaussian profile as used when fitting the much stronger [N~{\sc ii}~] emission at 6584 \AA. Most of the other features in the residuals of our stellar fit can be explained in light of the limitations of our adopted stellar templates (see Section~\ref{stacking:EMfitting}).  
For instance, the strong negative feature at 4668 \AA~corresponds to a C$_2$-absorption feature, which is not perfectly matched by the stellar model because the templates cannot account for the super-solar Carbon abundance that is observed in massive galaxies \citep{johansson12}. 

\begin{figure}
\centering
\includegraphics[clip=true,trim=0.8cm 0.5cm 0.8cm 0.5cm,scale=0.37,angle=90]{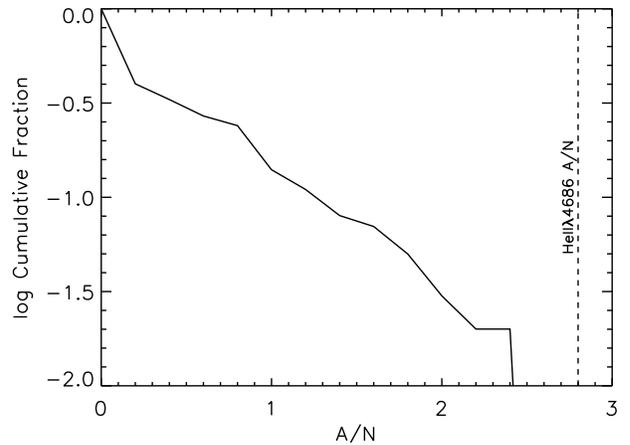}\\
\caption{The cumulative fraction of A/N values obtained when asking
  {\sc gandalf} to fit non-existing emission lines at 100
  randomly-generated spectral positions around the He~{\sc ii}~$\lambda$4686 line
  ($\pm$200 \AA). This gives
  the probability of finding a line with a specific A/N value, or
  above, just by chance. This test was run on the stacked
  spectrum obtained from our entire sample using the PCA criterion (11~593, see Section~\ref{data} and Table~\ref{tab:sample}), and while imposing on
  {\sc gandalf} to look for fake lines with the same Gaussian line
  profile as observed for the [N~{\sc ii}~]$\lambda$6548,$\lambda$6584 lines. We find similar probability distributions for all stacked spectra of the different age bins, as well as when repeating this experiment over
 different wavelength ranges (e.g. between 4800 -- 5200 \AA).}
  \label{AoN_distr}
\end{figure}

Table~\ref{AoN} lists the A/N values for the derived He~{\sc ii}~$\lambda$4686, H$\beta$ and [O~{\sc iii}~]$\lambda$5007 lines (cols. 3-5) for each stacked spectrum. In the comparison with modelled emission line predictions for the SD scenario we focus on emission line ratios instead of individual line fluxes (see Section~\ref{disc}). Hence, the value of the He~{\sc ii}~$\lambda$4686/H$\beta$ flux ratio is presented in Table~\ref{AoN} (col. 6), together with the [O~{\sc iii}~]$\lambda$5007/H$\beta$ line ratio (col. 7) used for constraining the ionization parameter (see Section~\ref{disc}). 
The strong [O~{\sc iii}~]$\lambda$5007 and H$\beta$ lines are always undoubtedly detected with A/N ratios greater than 40 and 23, respectively. The A/N ratios of the  He~{\sc ii}~$\lambda$4686 line instead falls in the range 2.6-3.0.
As mentioned above, the residual noise determining the A/N value of this weak line is dominated by systematics, i.e. the ability of the stellar templates to recover the stacked absorption spectra. 
 Thus, in the following section we assess the confidence level of the detection of the He~{\sc ii}~$\lambda$4686 line. 
At this point, we note that for all age bins we appear to find consistently low values (less than 0.1) for the He~{\sc ii}~$\lambda$4686/H$\beta$ ratio. We also note that the He~{\sc ii}~$\lambda$4686 flux appears to be slightly overestimated in the oldest age bin (see Appendix~A for more details).

\begin{figure}
\centering
\includegraphics[clip=true,trim=0.5cm 3.5cm 1.5cm 4.cm,scale=0.59,angle=90]{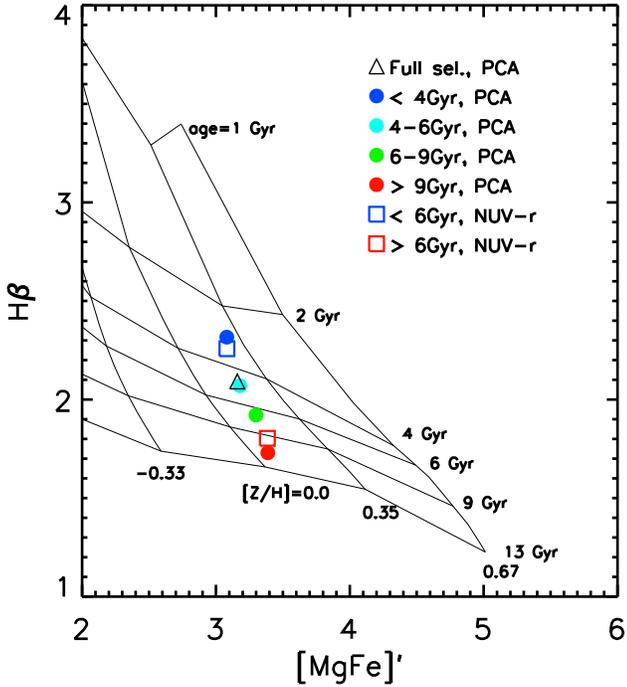}\\
\caption{Diagnostic diagram for the absorption lines indices H$\beta$
  as a function of [MgFe]' \citep{TMB03}, both known to be insensitive
  to [$\alpha$/Fe]-variations. Coloured points are indices derived on
  the emission line subtracted, stacked spectra 
  according to the labels. The black grids are stellar population
  models \citep{TMJ11} of varying age and metallicity as given by the
  labels and for [$\alpha$/Fe]=0.3. This shows that our division of
  individual galaxies into age-bins was successful.}
\label{HbMgFe}
\end{figure}

\begin{figure}
\centering
\includegraphics[clip=true,trim=0.3cm 0.5cm 0.3cm 1cm,scale=0.35,angle=90]{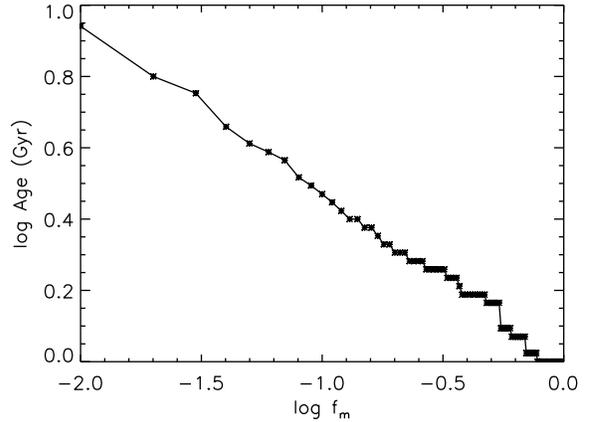}\\
\caption{The luminosity-weighted age of composite stellar populations consisting of the
mass fractions $f_\textrm{m}$ of a 1 Gyr old SSP and $1-f_\textrm{m}$ of a 10 Gyr old SSP. The
composite stellar populations were modelled using the spectra of SSP models from \citet{bc03} and the luminosity
weighted ages were derived using absorption line indices as described in Section~\ref{data:ages}.}
\label{BC03comp}
\end{figure}

\subsection{The A/N ratio as a tracer for the detection level of He~{\sc ii}~$\lambda$4686}
\label{results:AoN}

To further understand the connection between any measured value of the
A/N ratio and the level of detection for the He~{\sc ii}~$\lambda$4686 line, we have used
the {\sc gandalf} procedure to look for emission lines with the
same line profile of the strong [N~{\sc ii}~]$\lambda$$\lambda$6548,6584 lines
in spectral regions where no significant emission is known to exist. More
specifically, by repeating this experiment 100 times at random
spectral positions around the He~{\sc ii}~$\lambda$4686 line ($\pm$200 \AA) we
have assessed the probability of measuring any given A/N value just by
chance with our fitting procedure. In fact, assuming that the fit to
the stellar spectrum is not particularly worse around the position of the
He~{\sc ii}~$\lambda$4686 line than right on it, this test implicitly accounts for
spurious emission-line measurements (and corresponding A/N values)
that would be due to our limited ability to fit the real stellar
spectra of galaxies with our current set of stellar templates, 
resulting in {\sc gandalf} attempting to compensate for this problem
by means of a Gaussian feature.
The cumulative distribution for the fraction of A/N values returned by
our fake emission-line fits during this test is shown in
Fig.~\ref{AoN_distr}. The plotted distribution was derived from the
stacked spectrum corresponding to our entire sample using the PCA criterion (see Section~\ref{data:sf}), and similar
distributions are obtained also from the stacked spectra for each of
our stellar-age subsamples. 
This is not surprising as all spectra are characterized by similar systematic patterns of residuals around 4686~\AA\ (compare Fig.~\ref{fit:HaLT0p4} with Fig.~A1-A6). 
In the
He~{\sc ii}~$\lambda$4686 spectral region, the distribution plotted in
Fig.~\ref{AoN_distr} indicates that there is less than 5\% probability of
finding an A/N$\sim$2.0 or above, and less than 1\% probability for
values equal or in excess of an A/N$\sim$2.4.

The values of the A/N ratio for the derived He~{\sc ii}~$\lambda$4686 lines always
fall in the range between 2.6 and 3.0 (Table~\ref{AoN}). Given that
there is less than 1\% probability of measuring A/N values of 2.4 and above for the He~{\sc ii}~$\lambda$4686
line in our stacked spectra, this line is robustly detected  for all  groups of galaxies with different luminosity-weighted stellar ages.

\begin{figure*}
\centering
\includegraphics[clip=true,trim=0cm 0cm 9cm 19.3cm,scale=1,angle=0]{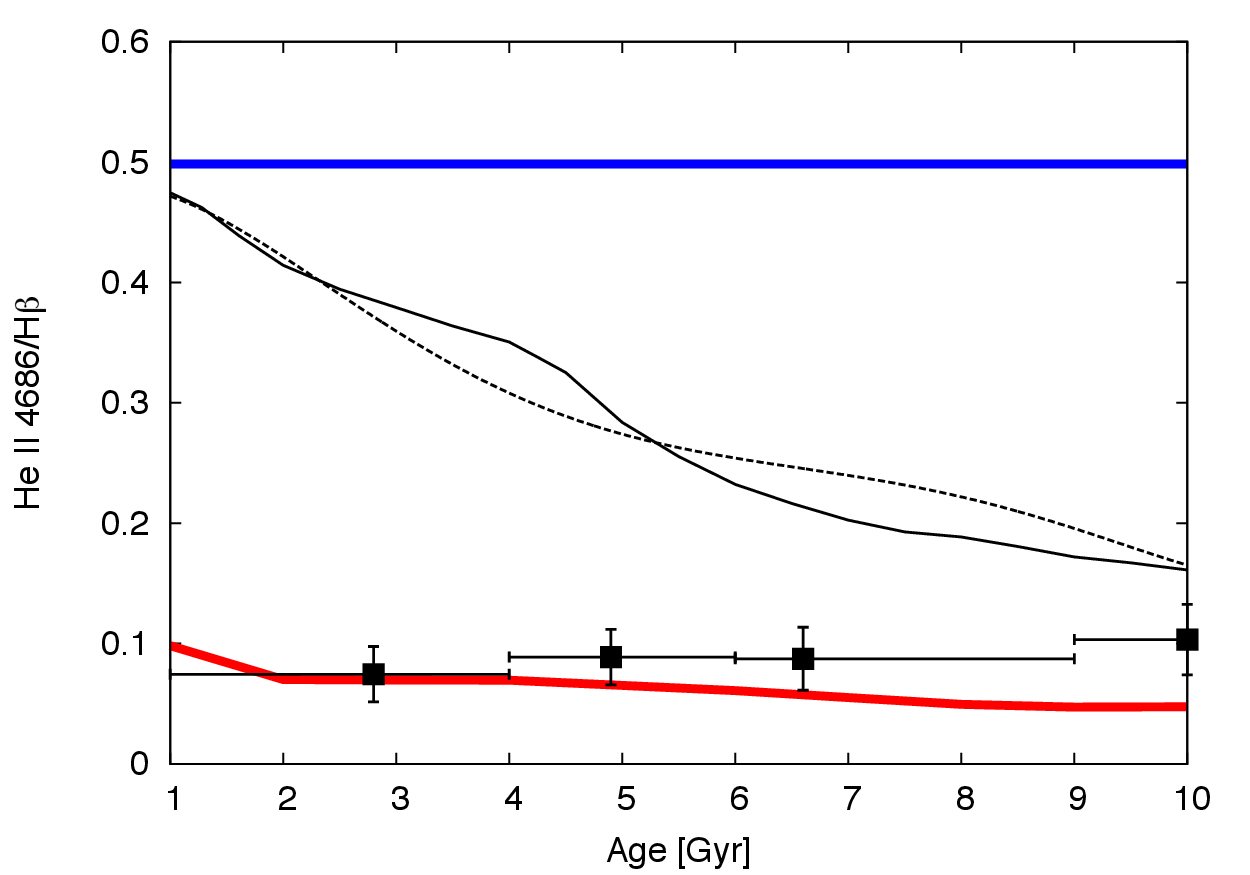}\\
\caption{ Comparison between observed values (filled squares, presented in Table~\ref{AoN}) of the
  He~{\sc ii}~$\lambda$4686/H$\beta$ line ratio and our model predictions as a function of stellar age.  
  The vertical and horizontal error-bars attached to our measurements represent the 1$\sigma$-errors on the line ratio and the range of stellar ages encompassed by each of our subsamples, respectively. The red and blue thick lines show the predictions for ionization by single type of ionizing sources, either by pAGB stars or by accreting WDs in the SD scenario for SN~Ia progenitors, respectively. The black solid line shows the predicted values for the He~{\sc ii}~$\lambda$4686/H$\beta$ ratio when both components are considered. In this case the spectrum of the stellar component (dominated by pAGB starts) was
computed using the SSP models from \citet{bc03}, whereas the luminosity of the SD SN~Ia
progenitors was computed using Eq.~1. In the latter, we assumed $\Delta\textrm{M}_\textrm{SB}=0.3\textrm{M}_\odot$, a WD effective temperature of $\textrm{T}_\textrm{eff}=2\cdot10^5$ K and the age
dependent SN~Ia rate as given by the delay time distribution of \citet{totani08}.
To illustrate the impact of non-SSP stellar populations, the dashed black line shows
the predictions for composite stellar populations combining 1 and 10 Gyr old SSPs (see text for
details).
  Both single-age and two-burst models predict declining He~{\sc ii}~$\lambda$4686/H$\beta$ values, reflecting the cosmic decrease in the SN~Ia rate. The observed values for the He~{\sc ii}~$\lambda$4686/H$\beta$ ratio are well below these models, appearing instead consistent with pAGBs being the only, or at least dominant, source of ionizing photons for the diffuse gas of quiescent galaxies.}
\label{fig:HeIIcomp}
\end{figure*}

\subsection{Stellar population ages}
\label{results:ages}

By subtracting the emission line spectra from the stacked galaxy spectra, we get clean absorption spectra for which we  derive the Lick indices (see Section~\ref{data:ages}). Fig.~\ref{HbMgFe} shows the classic H$\beta$ versus
[MgFe]' line-strength diagram \citep{TMB03} for the Lick indices derived on the stacked spectra, and with the TMJ models overlaid with ages and metallicities as given by the labels. This diagram is useful for visualizing stellar population parameters and the trend of the stacked spectra clearly follows that of the selected age bins. Furthermore, we apply our method for deriving stellar population parameters from observed Lick indices \citep[][see Section~\ref{data:ages}]{johansson12} and the derived ages for each of the stacks are presented in Table~\ref{AoN} (col.~8).  
These measurements confirm that the division into age bins, based on the individual spectra, was
successful. 

Even though we have excluded objects with
significant fractions of stars younger than a Gyr, our
luminosity-weighted ages could still be biased 
by the presence of a small fraction of $\sim$1 to 2-Gyr-old populations. 
To account for this possibility, we compute the luminosity-weighted age of a composite
stellar population consisting of the mass fraction $f_\textrm{m}$ of a 1 Gyr
old simple stellar population (SSP, i.e. single age and metallicity) superimposed on the fraction $1-f_\textrm{m}$ of a 10 Gyr old SSP. 
We adopt the SSP models from \citet{bc03}, to be consistent with the model predictions of \citet{woods}, 
with metallicities
of 2.5Z$_\odot$ and Z$_\odot$ for the young and old components, respectively. 
Instead using the models from \citet{mastro} gives a very similar result.
The
luminosity-weighted ages of the composite populations are determined from absorption line
indices as described in Section~\ref{data:ages}. The resulting dependence of the
luminosity-weighted age on the mass fraction $f_\textrm{m}$ of the 1 Gyr old SSP is shown in Fig.~\ref{BC03comp}. 
This dependence will be used in the next section to derive predictions for the SD scenario for the different composite populations to estimate the range of possible strengths of the He~{\sc ii}~$\lambda$4686 line at a given stellar population age.

%% file: Sec5.tex
\begin{figure*}
\centering
\includegraphics[clip=true,trim=0cm 0cm 9cm 19.3cm,scale=1,angle=0]{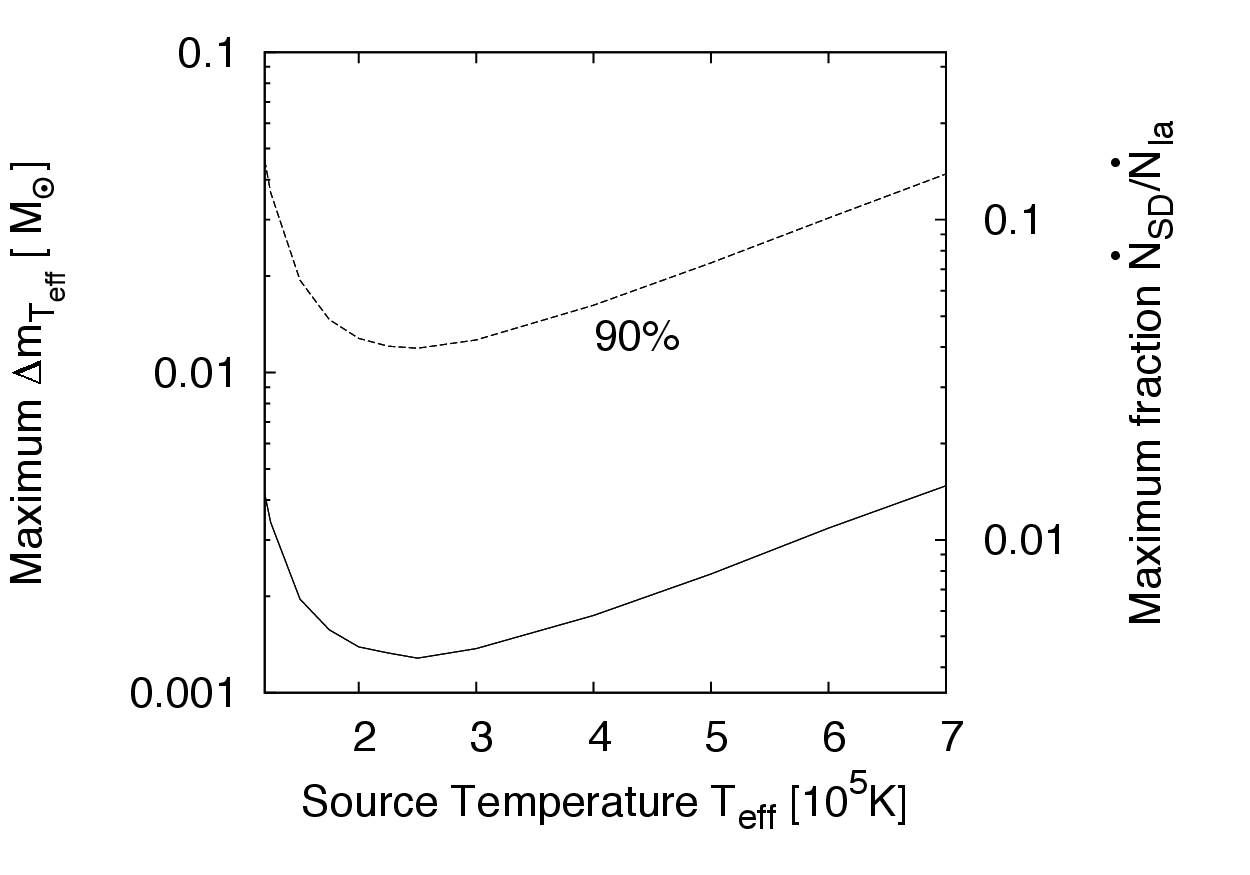}\\
\caption{The maximum mass accreted by a typical white dwarf -- successful SN~Ia progenitor -- as
a function of its assumed effective temperature for the delay time of $\approx 3$
Gyr, assuming that each SN~Ia progenitor spends some time in the stable nuclear
burning regime. The solid line shows the upper limit derived from the He~{\sc ii}~$\lambda
4686/H\beta$ ratio measured in the co-added spectrum for the youngest galaxy
sub-sample (1-4 Gyr). The dashed curve includes the statistical uncertainty in this ratio
at the 90\% confidence level. The right hand y-axis shows the  maximum fraction of
SN~Ia which can be produced in the SD scenario, assuming that each SD-progenitor
needs to accrete $\Delta\textrm{M}_\textrm{SB}=0.3 M_\odot$ of material.}
\label{fig:SDfrac}
\end{figure*}

\section{Constraints on the SD scenario}
\label{disc}

We now compare our emission-line results with the predictions of the photoionization models from \citet[][we refer the reader to this paper for details of the models]{woods} 
that include possible SD progenitors of SN~Ia. 
In particular, we focus on predicted values of the He~{\sc ii}~$\lambda$4686/H$\beta$ line ratio. This quantity remains as sensitive to the hardness of ionizing radiation as the He~{\sc ii}~$\lambda$4686 line itself, but  does not depend on assumptions regarding dust extinction or the fraction of ionizing photons that is intercepted by the interstellar medium. It does depend on the ratio of the ionizing photon flux to the gas density in units of the speed of light (per Hydrogen
atom, the so-called ionization-parameter $U$). This is constrained by the observed [O~{\sc iii}~]$\lambda$5007/H$\beta$ ratio (see Table~\ref{AoN}) following \citet{stasinska08} and \citet{yan12}. We find that in all age bins, the ratio [O~{\sc iii}~]$\lambda$5007/H$\beta$ is consistent with ionization by pAGB stars alone, with log(U) $\approx$ -3.5 (see Johansson et al., in preparation). 
This then fixes the incident ionizing photon flux at the face of the
nebula in our models (for a given mean density). Recall that the total
luminosity of any SD progenitor population is, in an average sense, fixed
by the SN Ia rate ($\dot{N}_{SNIa}$) and the total amount of mass processed through steady burning ($\Delta$M$_{SB}$, c.f. Eq. 1). Therefore,
varying the luminosity of the SD population through $\Delta$M$_{SB}$ effectively varies the hardness of the ionizing
radiation.

As a conservative estimate, we take $\Delta$M$_\textrm{SB}$ in Eq.~1 to be equal to 0.3 M$_\odot$, which corresponds to the difference of the critical Chandrasekhar mass ($\approx$1.4 M$_\odot$) and the maximum birth mass of a carbon-oxygen WD \citep[$\sim$1.1 M$_\odot$,][]{umeda99}. The observed He~{\sc ii}~$\lambda$4686/H$\beta$-ratio as a function of stellar population age is compared with the results of photoionization calculations for SSPs and the SD scenario in Fig.~\ref{fig:HeIIcomp}. For the latter, we take a characteristic temperature of $T_\textrm{eff}=2\cdot10^5K$, corresponding to the particular scenario of a white dwarf embedded in an optically thick radiation driven wind \citep{hachisu99}. Note that this photospheric temperature is also consistent with the lower range of temperatures estimated for supersoft sources \citep[e.g. SMC~3,][]{sturm11}. 
We produce photoionization models using both SSPs (solid line in Fig.~\ref{fig:HeIIcomp}) and composite populations (dashed line) for the ionizing emission. 
For the latter we use the fractions of 1~Gyr and 10~Gyr single burst populations derived in Section~\ref{results:ages} over the entire $1-10$~Gyr range. 
For luminosity-weighted ages above $\sim$6~Gyr, even a small fraction
of 1-Gyr-old stars in the composite models brings an overall higher
rate of SN~Ia and hence larger values of the predicted 
He~{\sc ii}~$\lambda$4686/H$\beta$ ratio than in the case of a pure old
stellar population, whereas the converse holds for lower values of the
luminosity-weighted age.  
The predictions shown in Fig.~\ref{fig:HeIIcomp} for the single-age case and
our particular composite models thus represent
conservative upper and lower boundaries for the
He~{\sc ii}~$\lambda$4686/H$\beta$ ratio that should bracket the case of less
extreme, and more plausible, star-formation histories.
The observed ratios are clearly well below those expected from models including a large population of accreting WDs. Notably, they are in agreement (within 1$\sigma$) with pABG stars being the sole ionizing source in retired galaxies. In Fig.~\ref{fig:HeIIcomp} we only present the observed values for the PCA selection (see Section~\ref{data:sf}), which are in good agreement with the observed values for the NUV-$r$ selection (see Table~\ref{AoN}).

Assuming that all WD progenitors of SN~Ia go through a steady nuclear-burning phase, we can constrain the maximum amount of mass accreted at any photospheric temperatures above $\sim 10^5$ K\footnote{Note that in the  wind models of \citet{hachisu99}, photospheric temperatures below $10^5$ K require mass accretion rates in excess of $10^{-5}$ M$_\odot$/yr, which would result in a prohibitively high mass loss rate and low mass accumulation efficiency for SN Ia progenitors. They are therefore irrelevant in the context of SN Ia progenitors in retired galaxies, because of the mass conservation considerations.} by requiring that our models do not conflict with the observed He~{\sc ii}~$\lambda$4686/H$\beta$-ratio. This is shown in Fig.~\ref{fig:SDfrac} (left axis) as a function of T$_\textrm{eff}$, derived for the mean age ($\approx$3 Gyr, see Table~\ref{AoN}) and observed He~{\sc ii}~$\lambda$4686/H$\beta$-ratio of the youngest age-bin ($<$4 Gyr) where the SN~Ia rate is high. This is derived accounting for the contribution from pAGBs. Taking the observed He~{\sc ii}~$\lambda$4686/H$\beta$-ratio at face value, it may be concluded that the maximum mass accreted by a typical white dwarf -- successful SN~Ia progenitor -- at  photospheric temperatures of  $(1-7)\cdot 10^5$ K lies in the range $\sim(1-4)\cdot 10^{-3}$ M$_\odot$. Note that the upper end of the temperature range is in fact more tightly constrained by X-ray observations \citep{gilfanov}. Accounting for the statistical errors on the He~{\sc ii}~$\lambda$4686/H$\beta$-ratio at the 90 \% confidence level, we obtain an upper limit of $\sim$0.04 M$_\odot$, also falling far short of the minimum 0.3M$_\odot$ required to trigger a SN~Ia explosion. 

Alternatively, we can estimate the maximum fraction of SN~Ia originating from the SD-channel, again assuming that a white dwarf needs to accrete at least $\Delta$M$_\textrm{SB,min}$=0.3 M$_\odot$ in order to produce a supernova. This fraction is shown on the right-hand axis of Fig.~\ref{fig:SDfrac} and does not exceed  $\sim 0.01$ for the observed He~{\sc ii}~$\lambda$4686/H$\beta$-ratio. Accounting for the errors at the 90 \% confidence level, we find an upper limit on the maximum fraction of SN~Ia originating from the SD-channel of $<0.1$.  Note that the typical value of $\Delta$M$_\textrm{SB}$ in the SD scenario should be larger than the assumed $\Delta$M$_\textrm{SB,min}$=0.3 M$_\odot$ resulting in an even more constraining upper limit. 

%% file: Sec6.tex
\section{Conclusions}
\label{conc}

We study nebular emission in passively evolving (``retired") galaxies  to 
constrain their populations of hot accreting WDs in the context of the SD scenario of SNe~Ia.  
In the SD scenario, a WD accretes hydrogen-rich material from a companion star until sufficient mass has accumulated to trigger a thermonuclear explosion. 
During the accretion phase, the WD emits significant UV radiation, peaking in the extreme UV when the photosphere is significantly inflated (as in some SD models).  
The collective emission from a population of such objects will alter the ionization balance in the warm/cold component of the interstellar medium in retired galaxies where mainly pAGB stars are expected (weak) ionizing sources.
Hence, the population of accreting WDs can be constrained by the strength of He~{\sc ii} recombination lines. 

For this purpose, we select galaxies from the SDSS with detected emission lines and where the ionizing background is not powered by AGNs or young stars in star forming regions. Furthermore, we select galaxies with stellar population ages $>1$Gyr that have not experienced star formation episodes in the last Gyr. In total we select $\sim11 500$ galaxies and their spectra are co-added in four bins of stellar population age resulting in $>2000$ galaxies in each bin, producing high quality stacked spectra with S/N values exceeding 800. For each stacked spectrum, we fit the SED in a wide wavelength range, 3800-6800 \AA , where a number of emission lines reside including He~{\sc ii}~$\lambda$4686, using the code {\sc gandalf} \citep{sarzi06} and the observed stellar library MILES \citep{miles} as templates for the stellar continuum. In the fitting process, the kinematics and position of He~{\sc ii}~$\lambda$4686 are constrained by the strong [N~{\sc ii}~]$\lambda$$\lambda$6548,6584 line doublet.

We extensively test the detection level of emission lines in the wavelength region around 4686~\AA\ and conclude that we, for the first time, detect the He~{\sc ii}~$\lambda$4686 line in retired galaxies. The line strengths of the stacked spectra in the four age bins are compared to the photoionization models from \citet{woods} that include the ionizing contribution of accreting WDs in the SD scenario. 
For this comparison we focus on the He~{\sc ii}~$\lambda$4686/H$\beta$ line ratio, which is insensitive to dust reddening and the covering fraction of gas. 
We find that in all age bins the He~{\sc ii}~$\lambda$4686/H$\beta$ ratio is (much) smaller than that 
expected in the SD scenario. For galaxies in the youngest age bin (1Gyr $<$ t$_\textrm{galaxy}$ $<$
4 Gyr), where the SN~Ia rate is the highest for the considered range of stellar ages, the He~{\sc ii}~$\lambda$4686 line is more than 10 times 
weaker than predicted by the SD scenario. On the other hand, the 
observed  He~{\sc ii}~$\lambda$4686/H$\beta$ line ratio is consistent with pAGB stars being the sole
ionizing source in all age bins.

These results exclude the presence of any significant population of accreting WDs with photospheric temperatures in the $\sim(1.5-7)\cdot 10^5$ K range. Thus,
only a small fraction of less than $\sim 5-10\%$ of SN~Ia observed in retired
galaxies could originate according to the classical version of the SD scenario.
Furthermore, irrespectively of the particular scenario, should all WD progenitors of
SN~Ia go through the phase of steady nuclear burning with the photospheric
temperature range $\sim(1.5-7)\cdot 10^5$ K, they do not increase their mass by
more than a few times $10^{-2}~\textrm{M}_\odot$ in this regime.

In this paper we have mainly focused on the He~{\sc ii}~$\lambda$4686 line, although 
we detect a large number of optical emission lines in the stacked spectra.  A general discussion of observed optical lines in retired galaxies, as well as the inferred properties of the associated dust, will be presented in a subsequent paper (Johansson et al., in preparation).

%% file: appendixA.tex
\appendix
\onecolumn
\begin{centering}
\section{Fits for all spectra}
\label{appendixA}
\end{centering}

\begin{figure*}
\centering
\includegraphics[clip=true,trim=0cm 0cm 0.5cm 0cm,scale=0.58,angle=90]{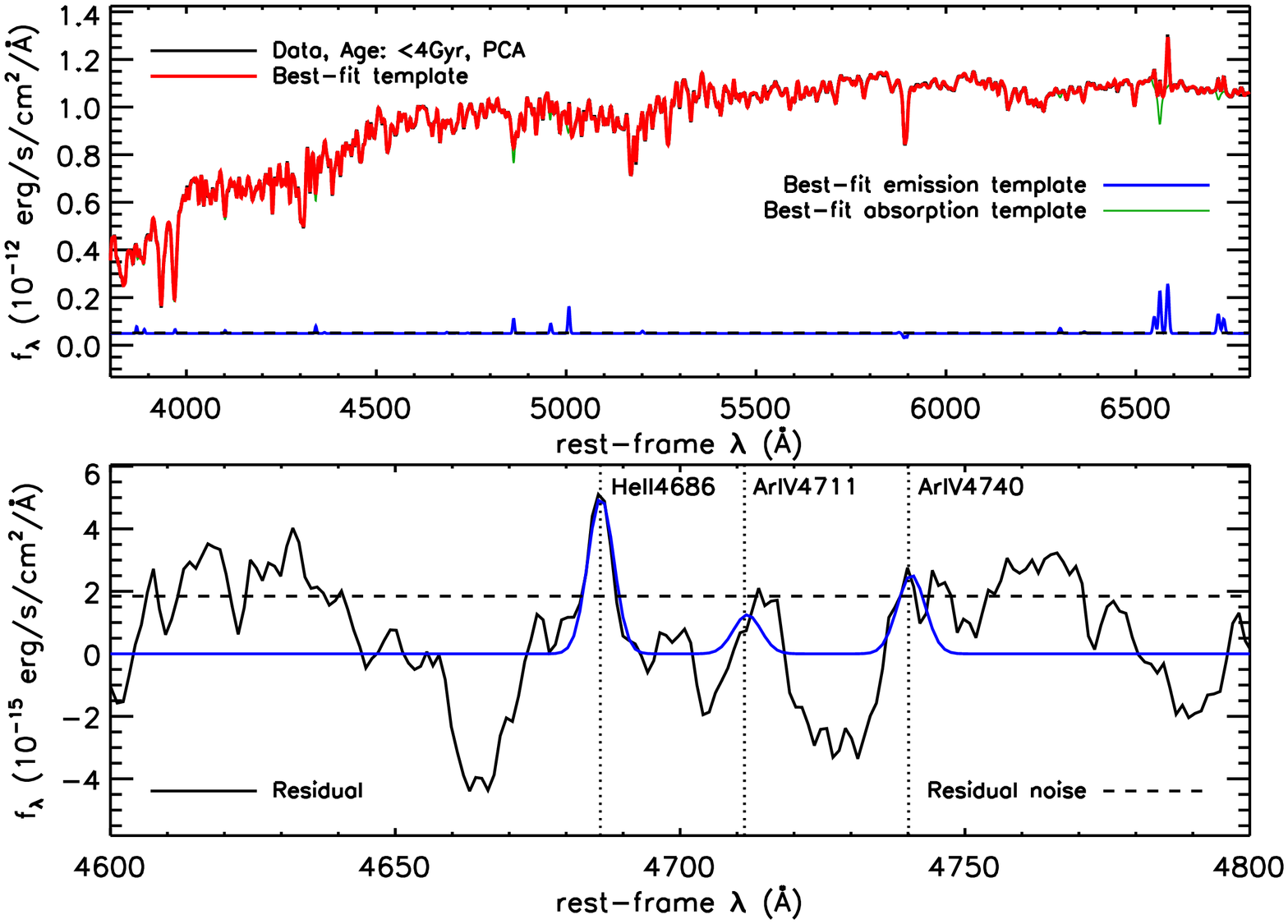}
\caption{Data and model of stacked spectrum. Upper panel: total spectrum obtained by co-adding the spectra
  of all galaxies in our sample with luminosity-weighted mean stellar ages $<$4 Gyr (2740
  objects) and selected using the PCA criterion (see Section~2), together with our best-fitting model (red line). The
  stellar component of this model is shown by the thin green line,
  whereas our best-fitting nebular spectrum is shown in blue. Lower
  panel: difference between the stacked spectrum and best-fitting
  stellar model around the position of the He~{\sc ii}~$\lambda$4686 line (black solid line),
  together with our best-fitting emission-line model (blue line). The
  dashed horizontal line shows the level of noise in the residuals
  (standard deviation) of the overall fit to the data in a wavelength region $\pm$200 \AA\ around 4686 \AA. In the
  4600-4800 \AA\ wavelength range, beside He~{\sc ii}~$\lambda$4686, we also fit
  for the [ArIV] lines at 4711 and 4740 \AA. The position of all lines
  is indicated by dotted vertical lines, with labels for their
  name. In this case, the ratio between the Gaussian model amplitude
  and the level of noise in the residuals, A/N, which relates to the
  detection level of the lines, is 2.7 for the He~{\sc ii}~$\lambda$4686-line.}
\label{fit:1}
\end{figure*}

\begin{figure*}
\centering
\includegraphics[clip=true,trim=0cm 0cm 0.5cm 0cm,scale=0.58,angle=90]{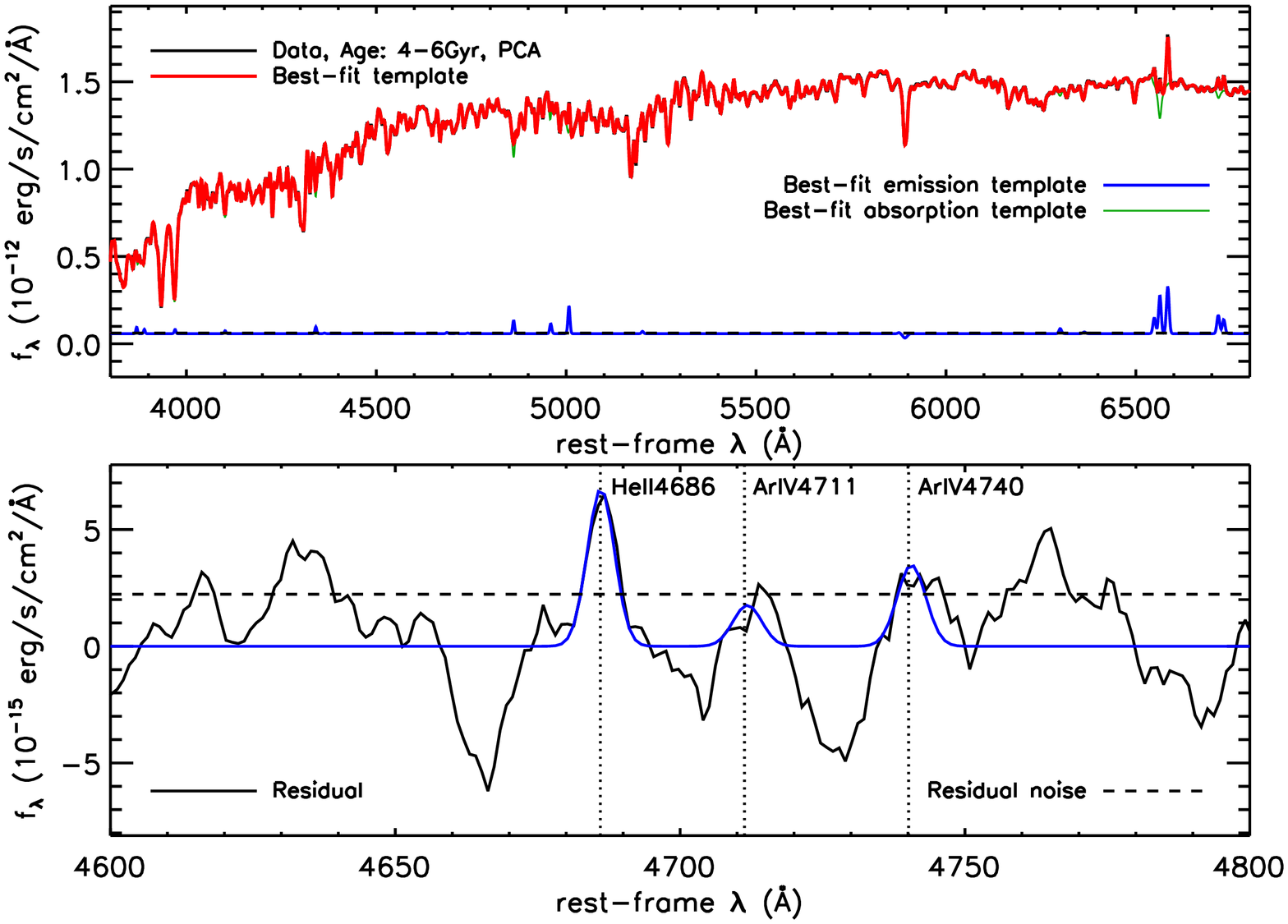}\\
\caption{Same as as Fig.~5 in the main manuscript and
  Fig.~\ref{fit:1}, but for the spectrum obtained by co-adding the
  spectra of all galaxies in our sample with luminosity-weighted mean
  stellar ages between 4 and 6 Gyr (3380 objects) and selected using the PCA criterion (see Section~2). The fit to the
  He~{\sc ii}~$\lambda$4686 line has an A/N=3.0.}
\label{fit:2}
\end{figure*}

\begin{figure*}
\centering
\includegraphics[clip=true,trim=0cm 0cm 0.5cm 0cm,scale=0.58,angle=90]{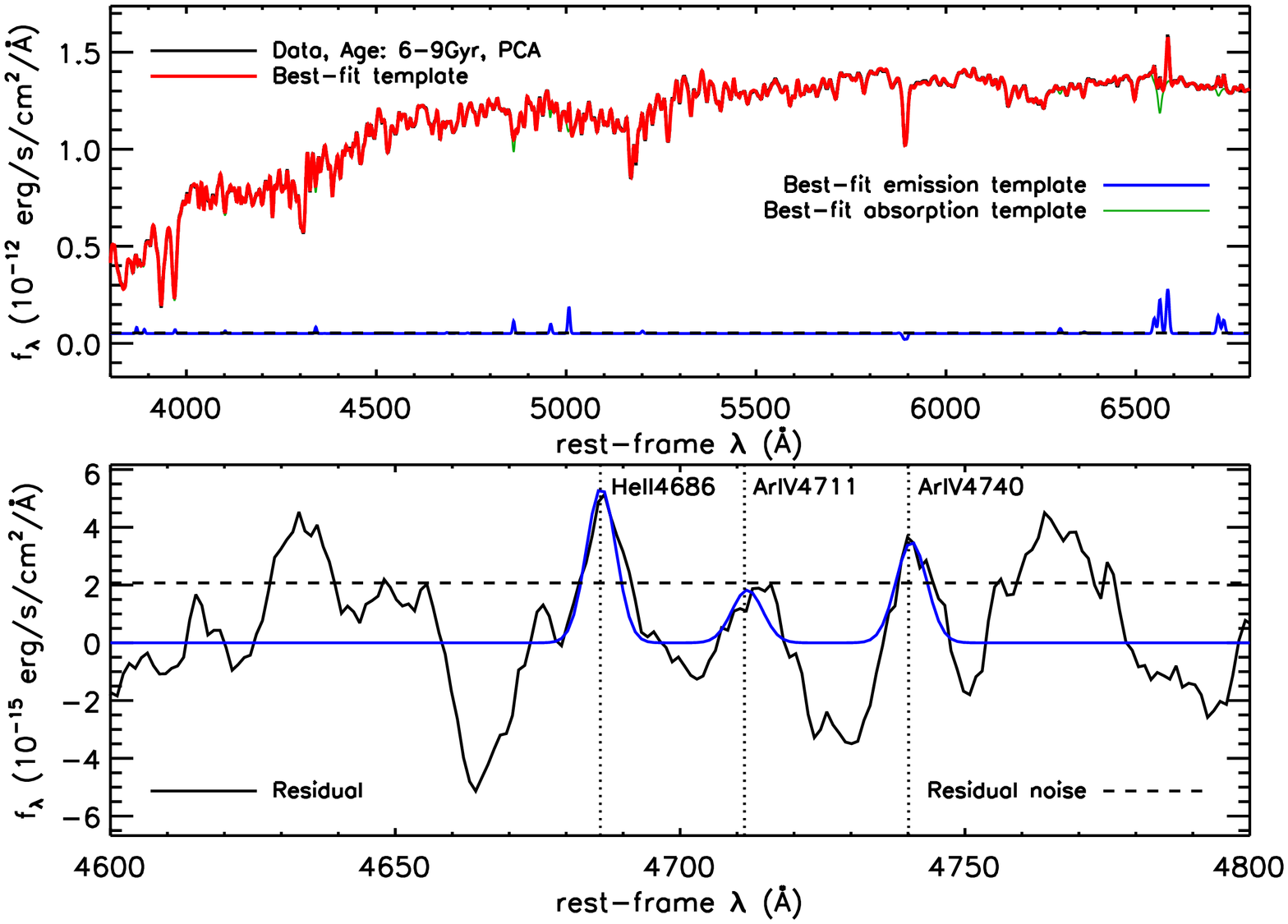}\\
\caption{Same as Fig.~5 in the main manuscript and
  Figs.~\ref{fit:1}-\ref{fit:2}, but for the spectrum obtained by
  co-adding the spectra of all galaxies in our sample with
  luminosity-weighted mean stellar ages between 6 and 9 Gyr
  (2729 objects) and selected using the PCA criterion (see Section~2). The fit to the He~{\sc ii}~$\lambda$4686 line has an
  A/N=2.6.}
\label{fit:3}
\end{figure*}

\begin{figure*}
\centering
\includegraphics[clip=true,trim=0cm 0cm 0.5cm 0cm,scale=0.58,angle=90]{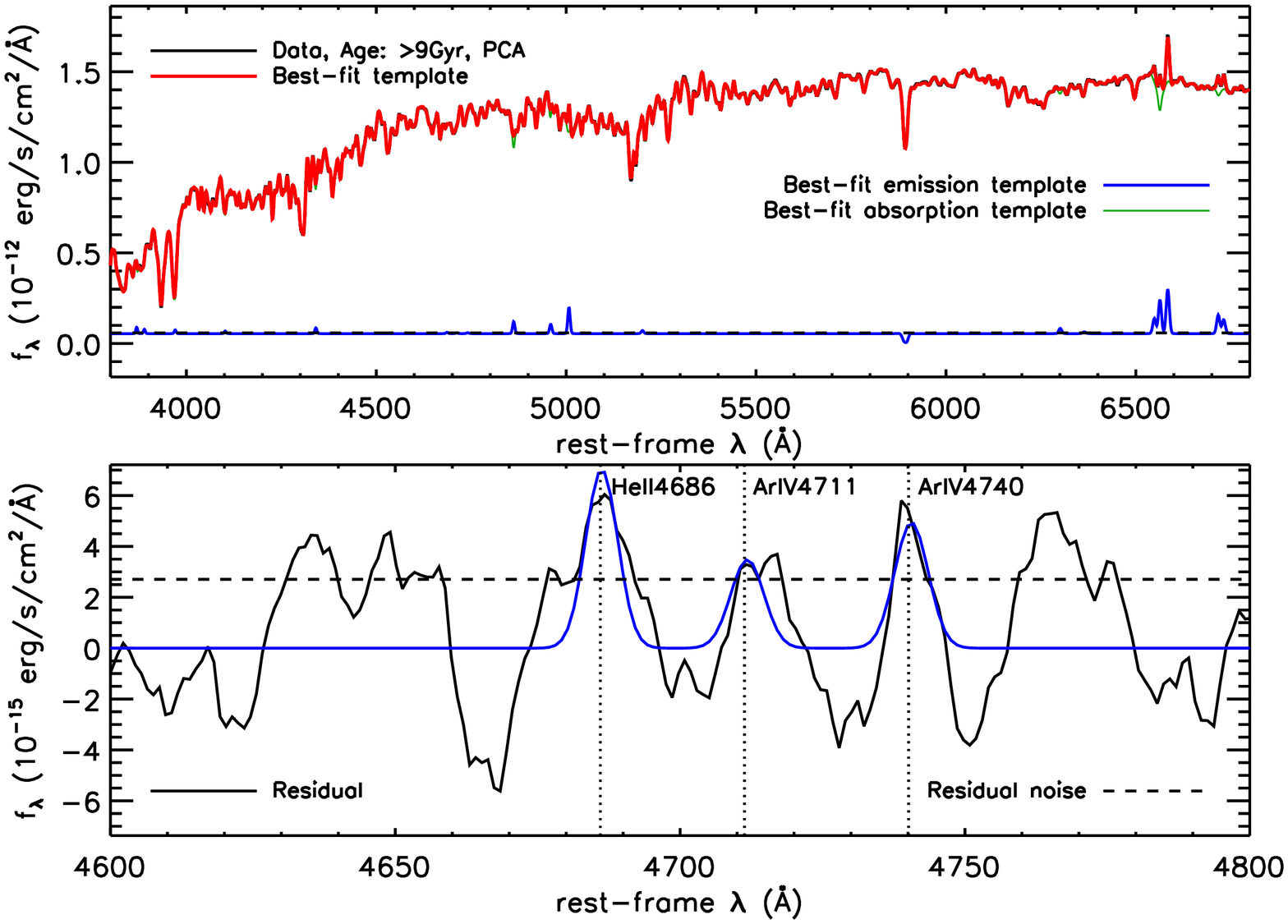}\\
\caption{Same as Fig.~5 in the main manuscript and
  Figs.~\ref{fit:1}-\ref{fit:3}, but for the spectrum obtained by
  co-adding the spectra of all galaxies in our sample with
  luminosity-weighted mean stellar ages $>$9 Gyr (2744 objects) and selected using the PCA criterion (see Section~2). The
  fit to the He~{\sc ii}~$\lambda$4686 line has an A/N=2.6.}
\label{fit:4}
\end{figure*}

\begin{figure*}
\centering
\includegraphics[clip=true,trim=0cm 0cm 0.5cm 0cm,scale=0.58,angle=90]{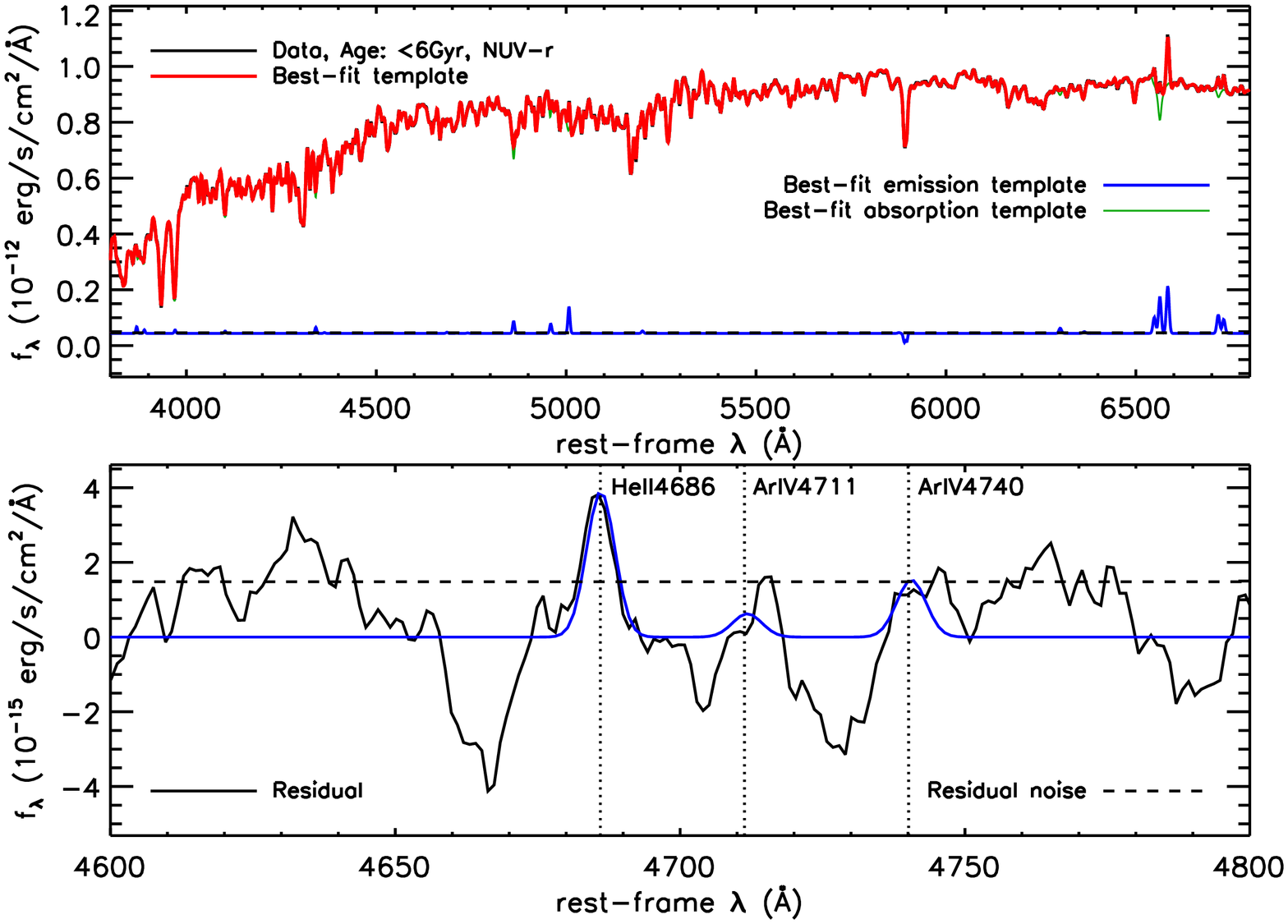}\\
\caption{Same as Fig.~5 in the main manuscript and
  Figs.~\ref{fit:1}-\ref{fit:4}, but for the spectrum obtained by
  co-adding the spectra of all galaxies in our sample with
  luminosity-weighted mean stellar ages $<$6 Gyr (1953 objects) and selected using the NUV-r criterion (see Section~2). The
  fit to the He~{\sc ii}~$\lambda$4686 line has an A/N=2.7.}
\label{fit:NUVr1}
\end{figure*}

\begin{figure*}
\centering
\includegraphics[clip=true,trim=0cm 0cm 0.5cm 0cm,scale=0.58,angle=90]{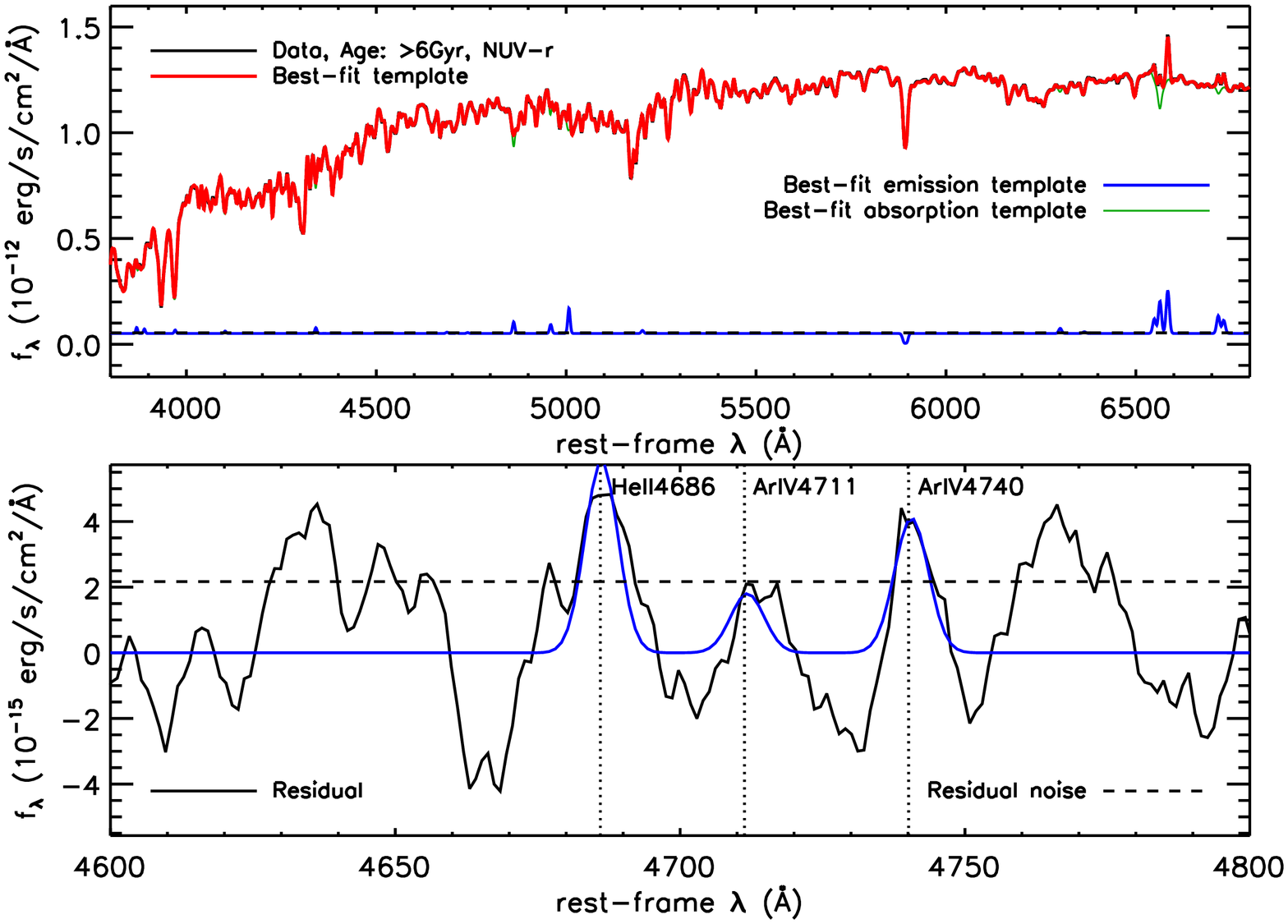}\\
\caption{Same as Fig.~5 in the main manuscript and
  Figs.~\ref{fit:1}-\ref{fit:NUVr1}, but for the spectrum obtained by
  co-adding the spectra of all galaxies in our sample with
  luminosity-weighted mean stellar ages $>$6 Gyr (2108 objects) and selected using the NUV-r criterion (see Section~5). The
  fit to the He~{\sc ii}~$\lambda$4686 line has an A/N=2.6.}
\label{fit:NUVr2}
\end{figure*}

\noindent Figs.~\ref{fit:1}-\ref{fit:NUVr2} show the stacked spectra for our four
age bins, together with their corresponding best-fitting stellar
model. Overall, we obtain very satisfactory fits also for the stacked
spectra of the subsamples. Furthermore, we nearly always find in the
residuals between the data and the stellar model a clear Gaussian
feature at the expected position of the He~{\sc ii}~$\lambda$4686 line, which
can be well matched using the same best-fitting line profile as that of the
strong [N~{\sc ii}~]$\lambda$6548,$\lambda$6584 lines. Only for the stacked
spectrum corresponding to our oldest age bins (Fig.~\ref{fit:4} and~\ref{fit:NUVr2} ), we
find a less regular and somewhat broader positive feature in the residuals between the data
and stellar model. Hence, the estimated He~{\sc ii}~$\lambda$4686
line flux of the oldest age bin should probably be regarded as an
upper limit.

The values of the A/N ratio for the He~{\sc ii}~$\lambda$4686 lines always
fall in the range between 2.6 and 3.0 (see Table~3). Given that
there is less than 1\% probability of measuring A/N values around 2.4 or above
just by chance in the wavelength region around the He~{\sc ii}~$\lambda$4686
line (see Section~4.2), this line is robustly detected also in the stacked
spectra for the groups of galaxies with different luminosity-weighted
stellar ages.

\begin{figure*}
\centering
\includegraphics[clip=true,trim=0cm 0cm 0.5cm 0cm,scale=0.58,angle=90]{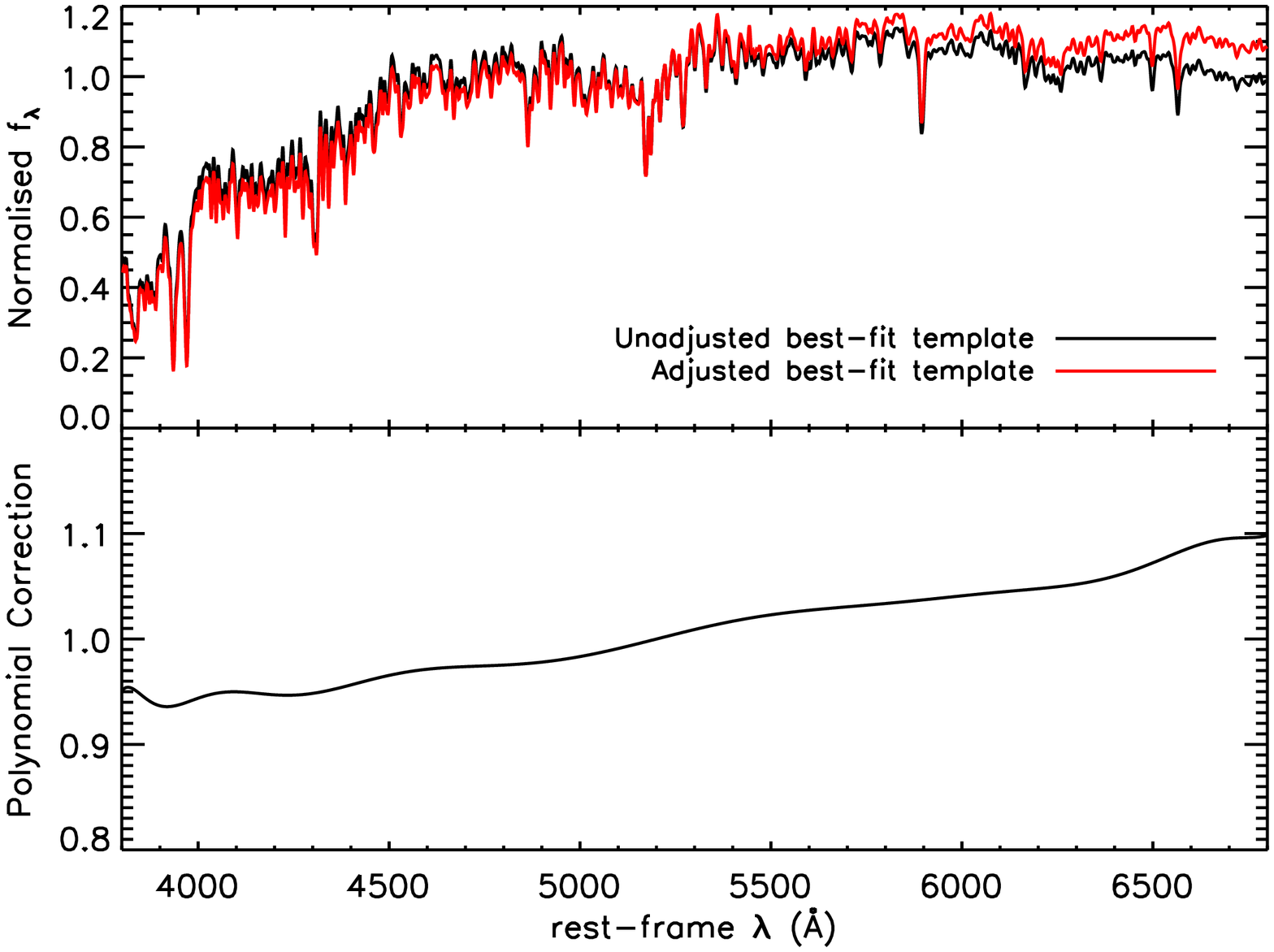}
\caption{Top panel: Best-fit template with (red spectrum) and without (black spectrum) polynomial correction (see Section~3.2), for the choice of stacking all individual spectra using the PCA criterion (see Section~2). Lower panel: Polynomial correction as a function of wavelength.}
\label{fit:correction}
\end{figure*}

To evaluate the fits in more detail, Fig.~\ref{fit:zoom} shows the fit around He~{\sc ii}~$\lambda$4686 (left hand side of each panel) and H$\beta$ and [O~{\sc iii}~] (right hand side of each panel) for each stacked spectrum, including best-fit overall template (red spectra), best-fit emission template (blue spectra) and best-fit stellar template (green spectra). We conclude that the quality of the fit is very high in these wavelength regimes. Moreover, to evaluate the effect of the polynomial correction (see Section~3.2), the top panel of Fig.~\ref{fit:correction} shows the best-fit template with (red spectrum) and without (black spectrum) polynomial correction, for the choice of stacking all individual spectra using the PCA criterion (see Section~2). The lower panel of Fig.~\ref{fit:correction} instead shows the polynomial correction as a function of wavelength, which shows a behavior close to being linear over the fitted wavelength range. Weak corrections on smaller scales are present, possibly due to flux calibration issues and template mismatch, but the polynomial correction is clearly dominated by the effects of reddening due to dust. Very similar results are found for all stacked spectra.

\begin{figure*}
\centering
\includegraphics[clip=true,trim=0cm 0cm 0.5cm 0cm,scale=0.35,angle=90]{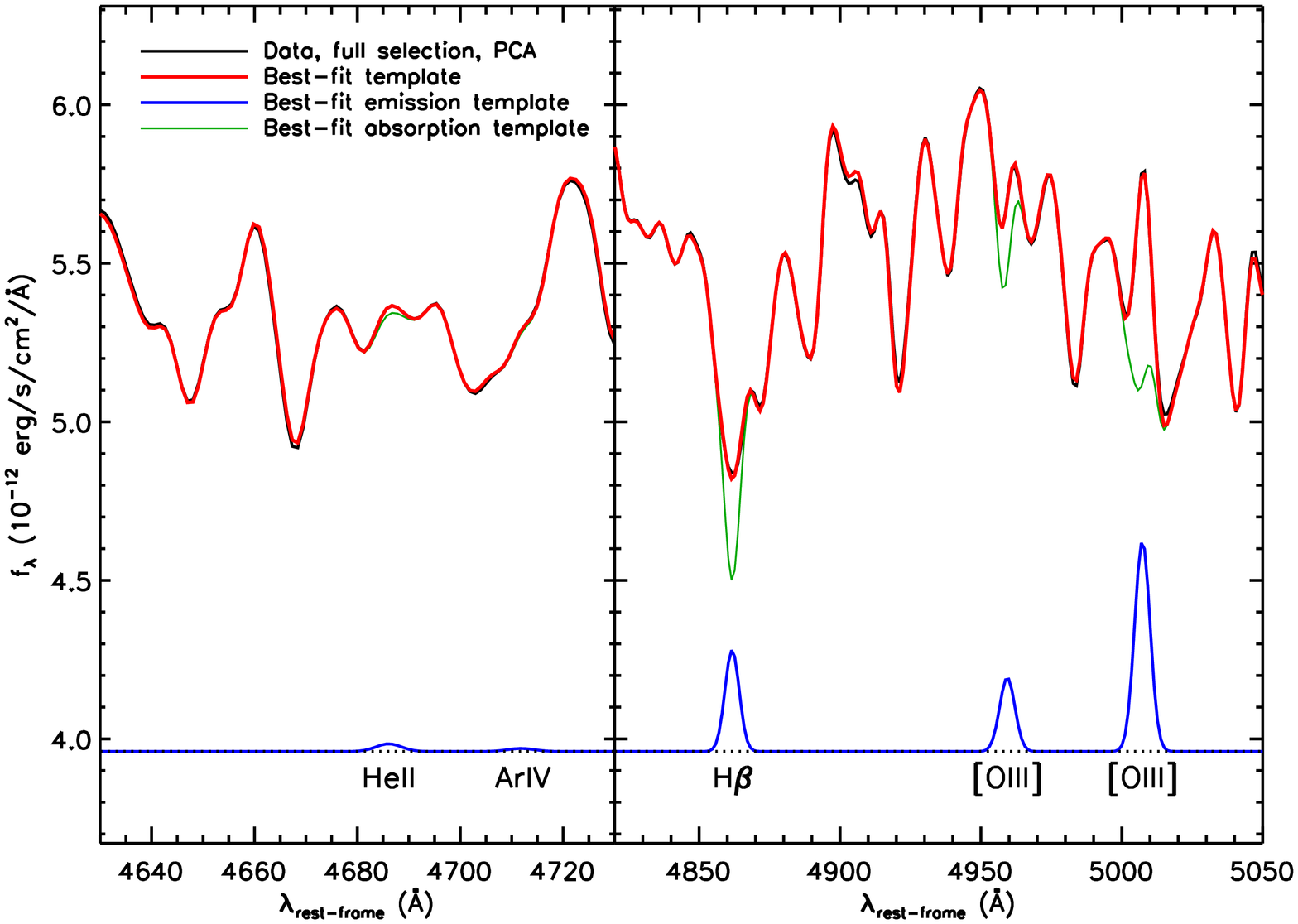}\includegraphics[clip=true,trim=0cm 0cm 0.5cm 0cm,scale=0.35,angle=90]{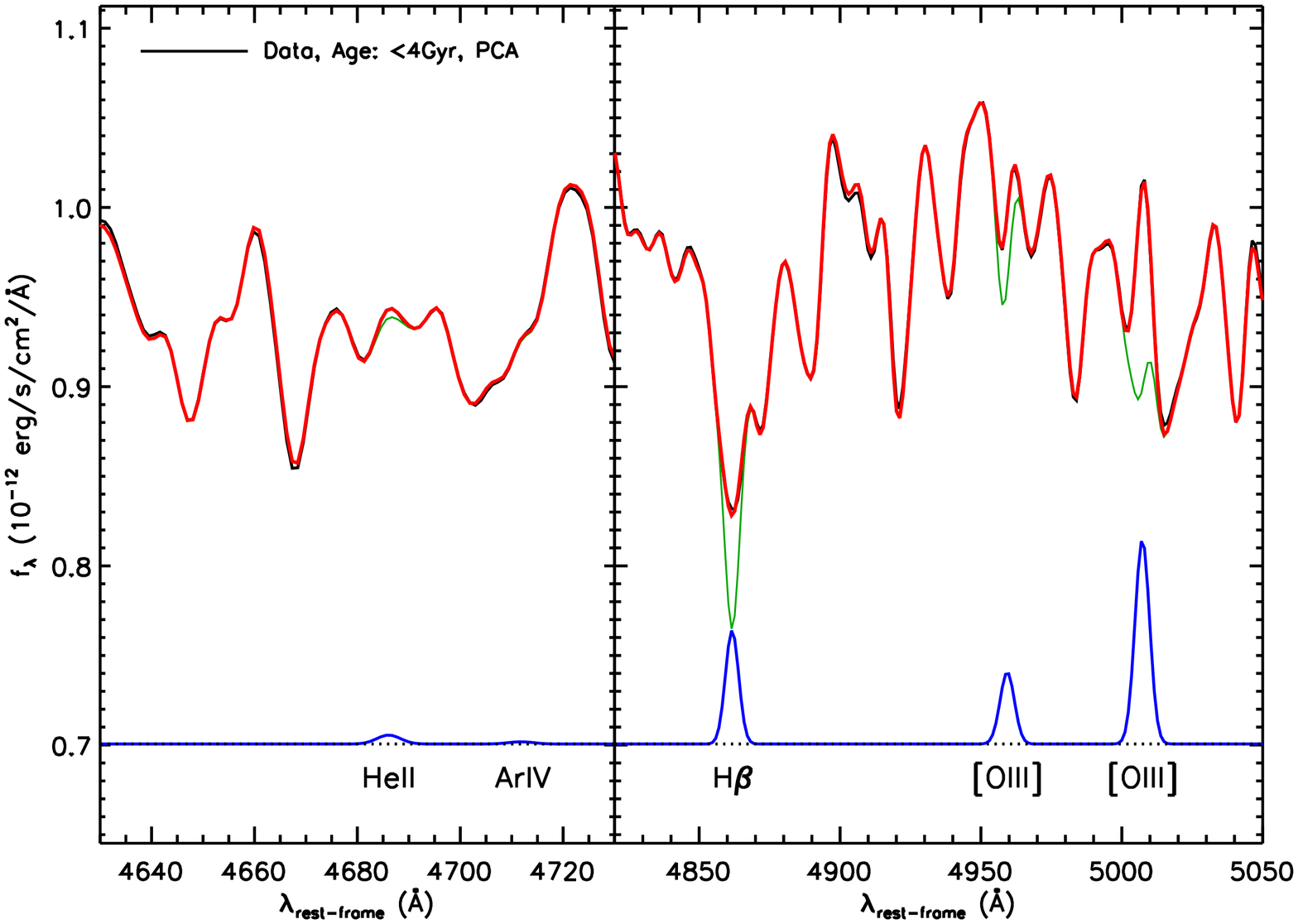}\\
\includegraphics[clip=true,trim=0cm 0cm 0.5cm 0cm,scale=0.35,angle=90]{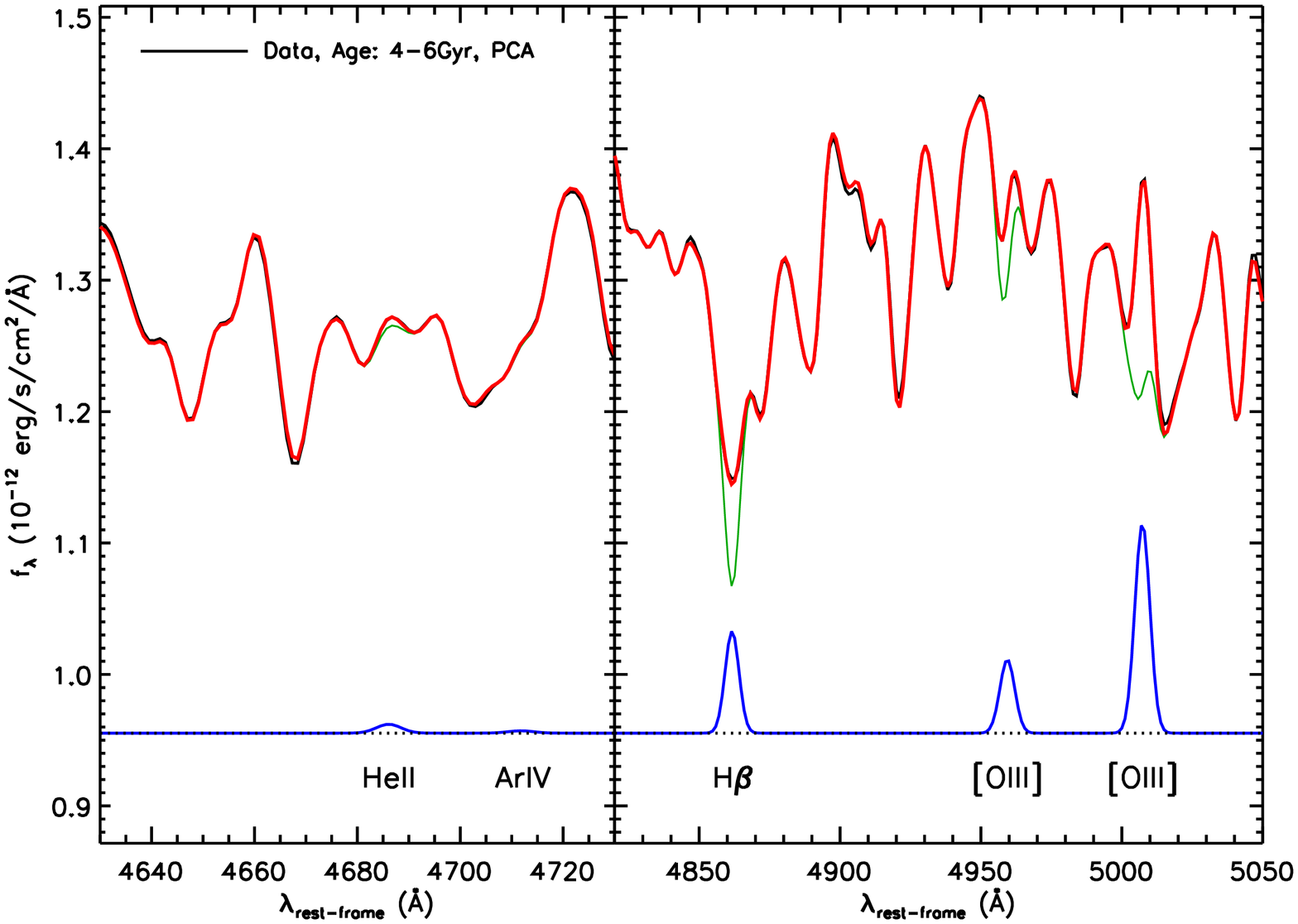}\includegraphics[clip=true,trim=0cm 0cm 0.5cm 0cm,scale=0.35,angle=90]{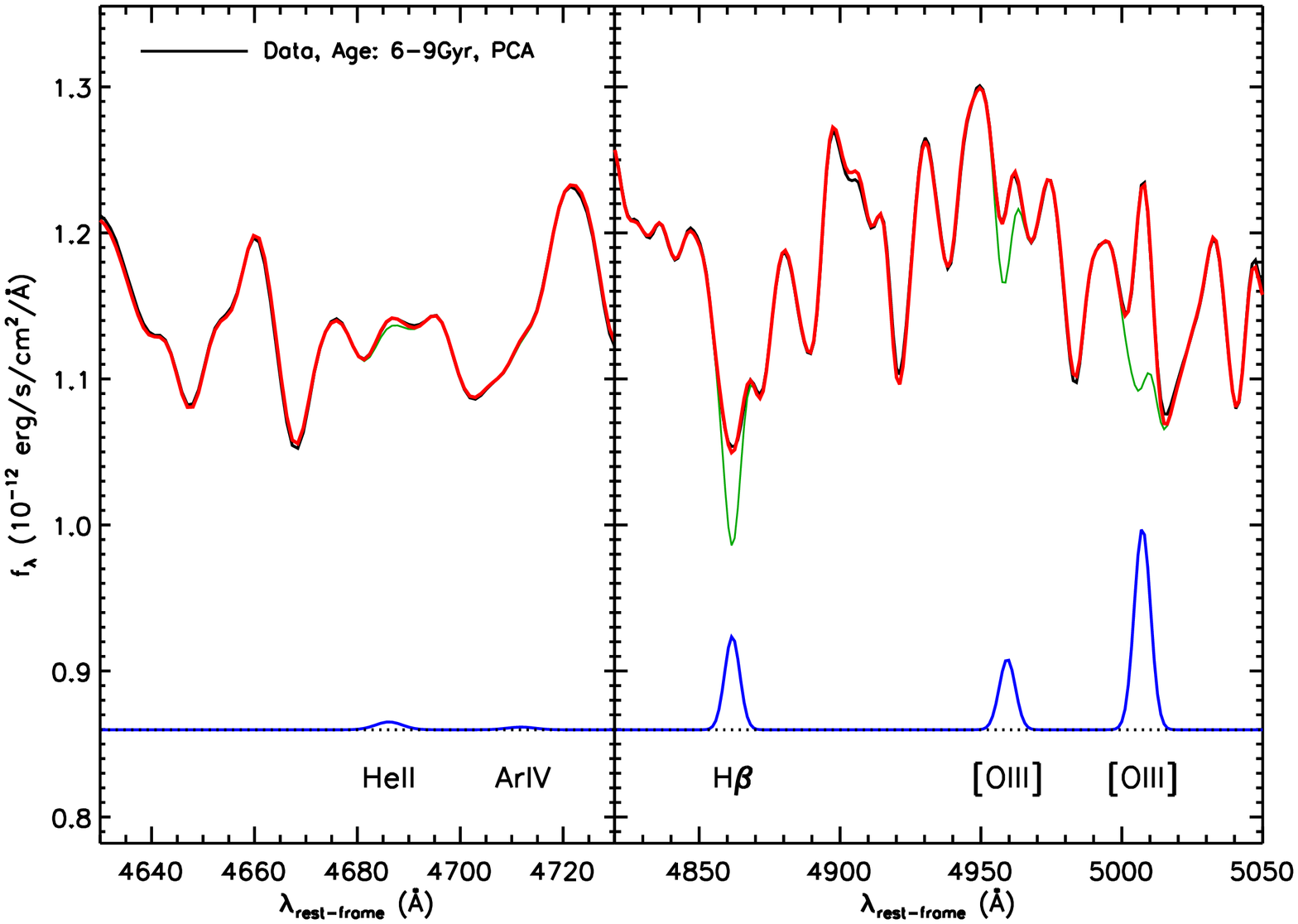}\\
\caption{Zoom-in on the fit around He~{\sc ii}~$\lambda$4686 (left hand side of each panel) and H$\beta$ and [O~{\sc iii}~] (right hand side of each panel) for each of the stacked spectra (see label in the upper left corner of each panel).}
\label{fit:zoom}
\end{figure*}

\begin{figure*}
\centering
\includegraphics[clip=true,trim=0cm 0cm 0.5cm 0cm,scale=0.35,angle=90]{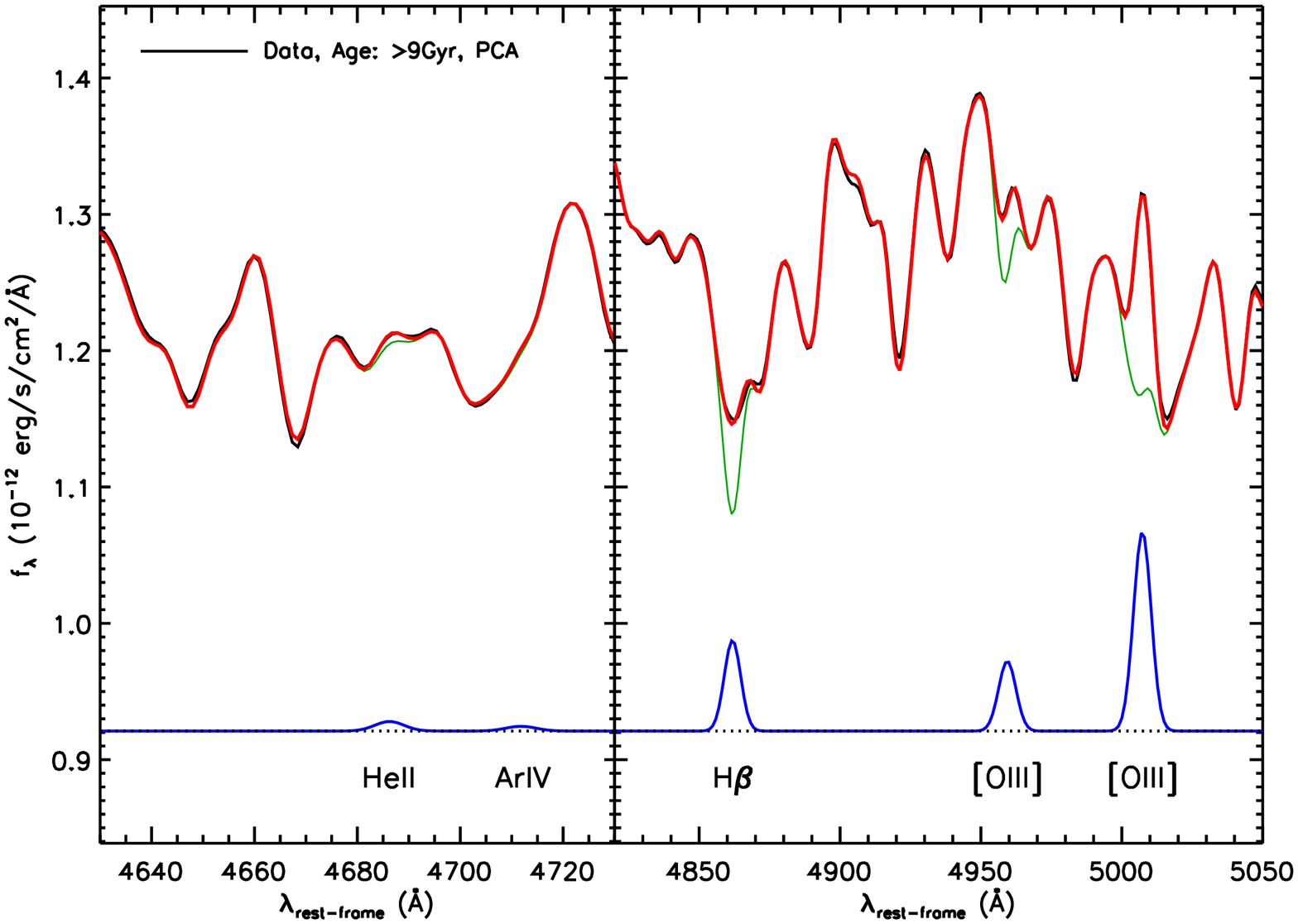}\includegraphics[clip=true,trim=0cm 0cm 0.5cm 0cm,scale=0.35,angle=90]{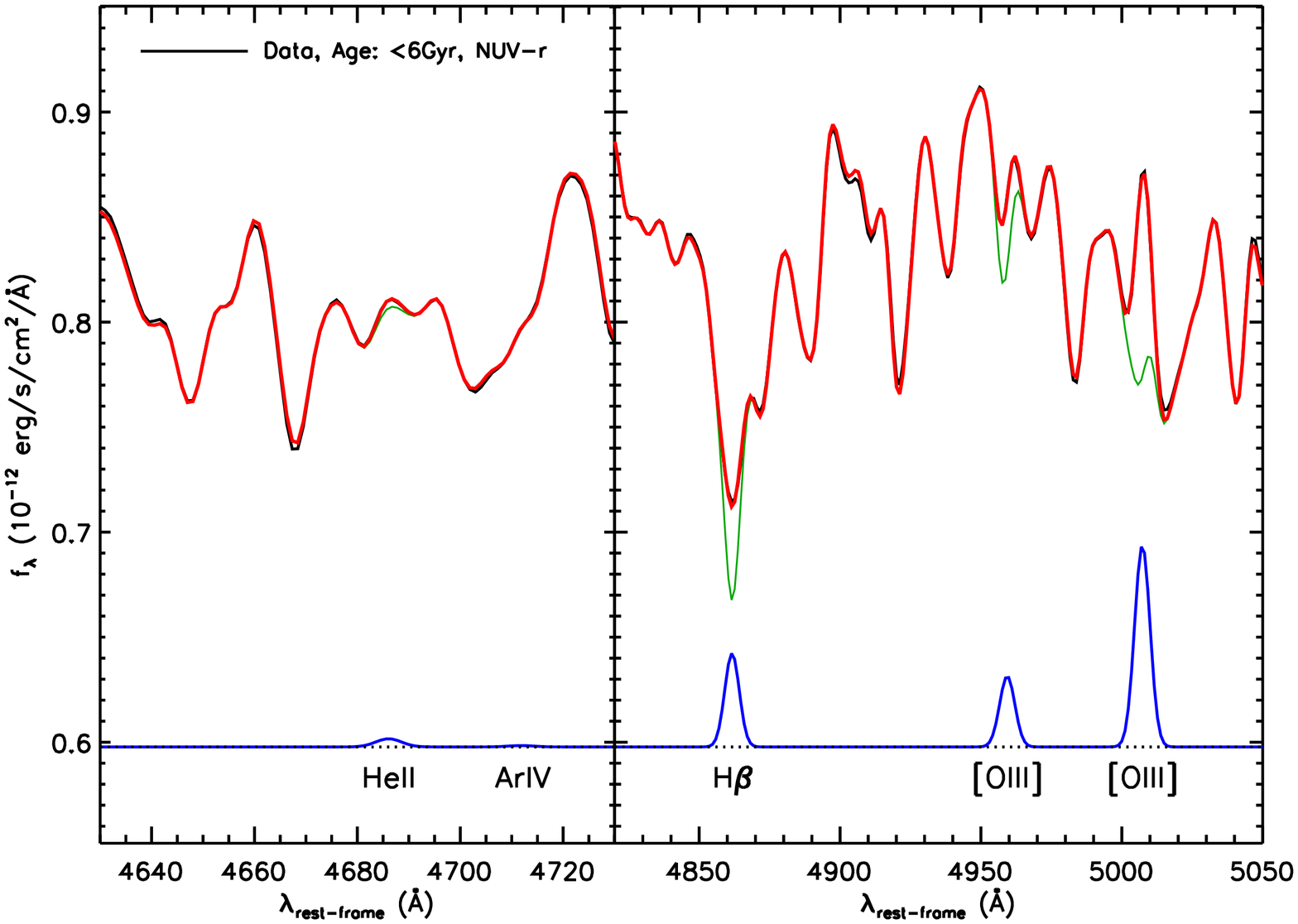}\\\includegraphics[clip=true,trim=0cm 0cm 0.5cm 0cm,scale=0.35,angle=90]{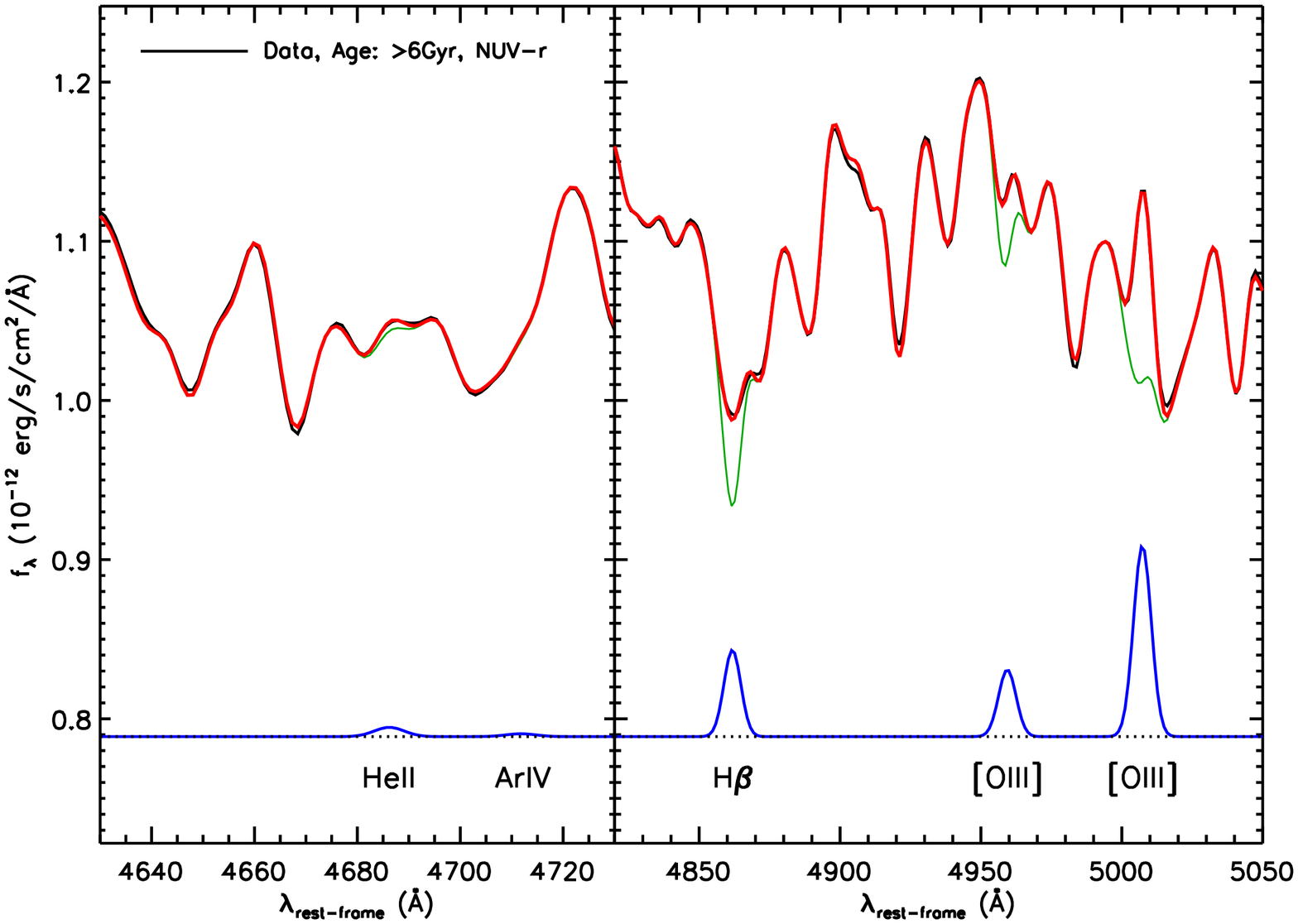}\\
\contcaption{}
\label{fit:zoom2}
\end{figure*}

